\documentclass[a4paper,11pt]{article}
\usepackage{graphicx}
\usepackage{color}
\usepackage{array}
\usepackage{amsmath}
\usepackage{amscd}
\usepackage{amsbsy}
\usepackage{amssymb}
\usepackage{theorem}
\usepackage{latexsym}
\usepackage{enumerate}
\usepackage{overcite}

\numberwithin{equation}{section}

\newcommand{\Z}{\ensuremath{\Vec{Z}}}

\makeatletter
\def\@cite#1{\textsuperscript{#1)}}
\makeatother

\title{\Large\textbf{
Exact Analysis of $\delta$-Function Attractive Fermions and Repulsive
Bosons in One Dimension}
}
\author{\large
Toshiaki \textsc{Iida}\thanks{iida@monet.phys.s.u-tokyo.ac.jp}\ and Miki  
\textsc{Wadati}\thanks{wadati@monet.phys.s.u-tokyo.ac.jp}
\\ \ \\
{\normalsize\it Department of Physics, Graduate School of Science, University of
Tokyo,}
\\{\normalsize\it Hongo 7-3-1, Bunkyo-ku, Tokyo 
113-0033, Japan}
}
\date{\small (March 2, 2005)}
\begin{document}
\pagenumbering{arabic}
\maketitle
\abstract{
The Gaudin integral equation for the ground state of a one-dimensional 
$\delta$-function attractive spin-$1/2$ fermions is solved in the form 
of power series. The first few terms of the asymptotic expansions for 
both strong and weak coupling cases are calculated analytically.   
The physical quantities such as the ground 
state energy are expressed in terms of a single dimensionless 
parameter $\gamma =c/D$, where $c$ is the coupling constant and $D$ is  
the number density. The results agree with those obtained from the 
perturbation calculations, which include the one in the classical 
electrostatics originally by Kirchhoff. In the strong coupling limit, 
the connection to the solutions of the Lieb-Liniger integral equation 
for the ground state of a one-dimensional $\delta$-function 
repulsive bose gas is shown explicitly.  
}\\

\begin{flushleft}
\textsf{KEYWORDS: $\delta$-function gas, one-dimensional integrable
 systems, BCS-BEC crossover, Lieb-Liniger integral equation, Gaudin
 integral equation, Tonks-Girardeau gas, hard-core bose gas} 
\end{flushleft}

\section{Introduction}\label{sec.intro}
We discuss a one-dimensional system of quantum $N$ particles with
$\delta$-function interaction in a periodic box of length $L$. The
Hamiltonian of the system is  
\begin{equation}
  H=-\sum^{N}_{j=1}\frac{\hbar^2}{2m}\frac{\partial^2}{\partial x_j^2}
  +c\sum_{i\neq j}\delta(x_i-x_j), 
\label{eq.hamiltonian}
\end{equation}
where $m$ is the mass and  $c$ is the coupling
constant. Hereafter, we choose a unit, $\hbar =1$ and $2m =1$ to
simplify the expressions. 

The system~\eqref{eq.hamiltonian} has been studied extensively by use of
the Bethe ansatz method~\cite{LL, Lieb, Gaudin67, Yang, Gaudin71,
Mattis, Gaudin, Kore, Takahashi}.  
For $N$ repulsive (i.e. $c > 0$) bosons, the ground state is
described by the integral equation for the distribution function
$\rho(k)$ of the quasi-momenta~\cite{LL}, 
\begin{gather}
 \rho(k)=\frac{1}{2\pi}+\frac{c}{\pi}\int_{-K}^{K}\frac{\rho 
  (q)\,dq}{c^2+(k-q)^2}, \label{eq.LL}
\end{gather} 
where $K$ is a cut-off momentum, which is fixed by the normalization
condition, 
\begin{equation}
 \int_{-K}^{K}\rho(k)\,dk = D.
\end{equation} 
Here, $D\equiv N/L$ is the number density of the particles. 
The interval $[-K,K]$ is filled with
quasi-momenta $k$. The ground state energy is given by
\begin{equation}
  \frac{E_\mathrm{B}}{L}=\int_{-K}^{K}k^2\rho(k)\,dk.\label{eq.bose_energy}   
\end{equation}  
We call~\eqref{eq.LL}
the Lieb-Liniger integral equation.   

In 1967, Yang~\cite{Yang} proposed a method, which we now call the
generalized Bethe ansatz, or ``nested'' Bethe ansatz, to study the 
system~\eqref{eq.hamiltonian} with arbitrary internal degree of freedom.
For $N$ attractive (i.e. $c<0$) spin-$1/2$ fermions, the 
ground state with total spin $0$ is described by~\cite{Gaudin67}
\begin{equation}
 f(q)=2+\frac{c}{\pi}\int_{-Q}^{Q}\frac{f(q')\,dq'}
 {c^2+(q-q')^2}. \label{eq.GY}
\end{equation}
Here, $f(q)$ is the distribution of spins with spin-wave rapidity 
$q$, and a cut-off rapidity $Q$ is fixed by the normalization condition for
$f(q)$:
\begin{equation}
 \frac{1}{\pi}\int_{-Q}^{Q}f(q)\,dq=D, \label{eq.normalization_GY}
\end{equation}
while the ground state energy is given by 
\begin{equation}
  \frac{E_{\mathrm{F}}}{L}=\frac{1}{\pi}
 \int_{-Q}^{Q}\left(q^2-\frac{c^2}{4}\right)f(q)\,dq. 
\label{eq.fermi_energy}
\end{equation}
For our purpose, it is convenient to introduce the effective ground state
energy,
\begin{equation}
 \frac{E_{\mathrm{F}}^{\mathrm{eff}}}{L}\equiv
  \frac{E_{\mathrm{F}}}{L}+\frac{c^2}{4}D
  =\frac{1}{\pi}
  \int_{-Q}^{Q}q^2f(q)\,dq. \label{eq.effective_fermi_energy}
\end{equation}
We call~\eqref{eq.GY} the Gaudin integral equation. 

By use of a method proposed in the previous paper~\cite{MW02} to
solve~\eqref{eq.LL}, we explicitly calculate the first few terms of the
asymptotic expansions of $f(q)$ for both strong and weak coupling cases.  
It is essential to use the single dimensionless coupling
constant $\gamma$, defined by
\begin{equation}
 \gamma=\frac{c}{D}=\frac{2mc}{\hbar^2 D}. \label{eq.def_gamma}
\end{equation}
Combining the results with those in  the previous papers~\cite{MW02,TI},
we discuss in particular four cases, $\gamma\to -\infty$, $\gamma \to
-0$, $\gamma \to +0$, and $\gamma\to +\infty$. We also study the physical  
quantities, such as the ground state energy and the chemical potential,
as functions of $\gamma$.   

This paper is organized as follows. In \S~\ref{sec.solutions}, we 
solve~\eqref{eq.GY} for large and small $|\gamma|$. The similarities   
and the connections of the obtained expressions in the case of
$\gamma\to -\infty$ for~\eqref{eq.GY} and $\gamma\to +\infty$
for~\eqref{eq.LL} are investigated in \S~\ref{sec.BCS-BEC}. The  
recent discussion on ``BCS-BEC crossover''~\cite{Batchelor, Tokatly, 
Fuchs} is made clear for these integrable models in one-dimension. For
the small $|\gamma|$ case, it is known to be difficult to
solve~\eqref{eq.LL} and~\eqref{eq.GY}. Analogies with the perturbation
calculation in the classical electrostatics originally by Kirchhoff are
discussed in \S~\ref{sec.condenser}. The last section is devoted to the 
conclusions. To concentrate on physical discussions, most of the
calculations are given in
Appendices~\ref{ap.another_small_lambda}$\sim$\ref{app.pf_sum}.

\section{Solutions of the Gaudin Integral Equation}\label{sec.solutions}
\subsection{Formulation}
Following a method presented in the previous papers~\cite{MW02,TI}, we  
solve the Gaudin integral equation~\eqref{eq.GY}. In terms of new
variables,  
\begin{align}
q &= Qx, & c &= -Q\lambda, & q' &= Qy, \label{eq.variables} \\
& & f(Qx) &= F(x), & &  \label{eq.def_of_F}
\end{align}
\eqref{eq.GY} becomes
\begin{equation}
F(y)+\frac{\lambda}{\pi}\int^{1}_{-1}\frac{F(x)\,dx}{
\lambda^2+(x-y)^2}=2.\label{eq.dimless_GY}
\end{equation}
We also refer to~\eqref{eq.dimless_GY} as the Gaudin integral
equation. As mentioned, we use the number density $D$ and the
dimensionless coupling constant $\gamma$, 
\begin{gather}
 D=\frac{N}{L}, \label{eq.D} \\
\gamma=\frac{c}{D}.\label{eq.gamma}
\end{gather}
The large (small) $\lambda$ corresponds to the large (small) $|\gamma|$. 

By expanding
\begin{equation}
F(x)=\sum^{\infty}_{n=0}a_n x^{2n},\label{eq.expansion_F}
\end{equation}
we have
\begin{align}
&\int^{1}_{-1}dx\,\frac{F(x)}{\lambda^2+(x-y)^2} \nonumber\\
  &= \sum^{\infty}_{n=0}\sum^{\infty}_{p=0}\sum^{p}_{m=0}
	\frac{a_n\cdot 2^{2m+1}}{(2m)!\,(p-m)!}\nonumber\\
&\phantom{=} \times\frac{\partial^{m+p}}{\partial
 s^{m+p}}\left[\sum^{m+n}_{l=1}\frac{(-1)^{l+1} 
	\lambda^{2(l-1)}}{2(m+n-l)+1}+(-1)^{m+n}\lambda^{2(m+n)-1}
	\arctan{\frac{1}{\lambda}}\right]y^{2p},\label{eq.kernel}
\end{align}
where $s=\lambda^2$. Substituting \eqref{eq.expansion_F} and
\eqref{eq.kernel} into~\eqref{eq.dimless_GY}, we get
\begin{align}
 &\sum^{\infty}_{p=0} a_p y^{2p}+\frac{\lambda}{\pi}
 \sum^{\infty}_{n=0}\sum^{\infty}_{p=0}\sum^{p}_{m=0}
 \frac{a_n \cdot 2^{2m+1}}{(2m)!\,(p-m)!}\nonumber
 \\
 &\qquad\times\frac{\partial^{m+p}}{\partial s^{m+p}}
 \left[\sum^{m+n}_{l=1}\frac{(-1)^{l+1}\lambda^{2(l-1)}}
 {2(m+n-l)+1} + (-1)^{m+n}\lambda^{2(m+n)-1}\arctan{\frac{1}
 {\lambda}}\right] y^{2p}=2. \label{eq.exp_GY}
\end{align}
In order that~\eqref{eq.exp_GY} holds for any $y$,
$-1\le y\le 1$, we equate the terms of the same power of $y^{2p}$. At
$y^0$ and $y^{2p}$ $(p\ge 1)$, we respectively obtain

\begin{subequations}\label{eq.ap}
\begin{align}
\begin{split}
 a_0&+\frac{2}{\pi}\arctan{\frac{1}{\lambda}}\cdot a_0 \\
 &+ \frac{2\lambda}{\pi}\sum^{\infty}_{n=1}
 a_n \left(\sum^{n}_{l=1}\frac{(-1)^{l+1}\lambda^{2(l-1)}}
 {2(n-l) + 1} + (-1)^{n}\lambda^{2n-1}\arctan{\frac{1}{\lambda}}
 \right)= 2,\label{eq.ap_1}
\end{split}
& \\
 \begin{split}
  a_p&+\frac{2\lambda}{\pi}\sum^{\infty}_{n=p+1}\sum^{p}_{m=0}
  \frac{a_n \cdot 2^{2m}}{(2m)!\,(p-m)!}\frac{\partial^{m+p}}
  {\partial s^{m+p}} \left[\sum^{m+n}_{l=1}\frac{(-1)^{l+1}s^{l-1}}
  {2(m+n-l)+1}\right] \\
  &+ \frac{2\lambda}{\pi}\sum^{\infty}_{n=0}\sum^{p}_{m=0}
  \frac{a_n \cdot 2^{2m}}{(2m)!\,(p-m)!}\frac{\partial^{m+p}}
  {\partial s^{m+p}} \left[(-1)^{m+n} \lambda^{2(m+n)-1}
  \arctan{\frac{1}{\lambda}}\right]=0.\label{eq.ap_p}
 \end{split}
&
\end{align}
\end{subequations}
In this way, a solution of~\eqref{eq.ap} gives a solution 
$F(x)$ of~\eqref{eq.dimless_GY}.
 
Until now, all the expressions are exact and no approximation is
employed. We summarize here the method of solution: 
\begin{enumerate}
 \item Solve~\eqref{eq.ap} to determine $\{a_n\}$ of
       \eqref{eq.expansion_F} in the case of large $\lambda$ and small
       $\lambda$. 
 \item Use~\eqref{eq.normalization_GY} with~\eqref{eq.variables}, which
       can be rewritten as   
       \begin{equation}
	\gamma = -\frac{\pi\lambda}{\displaystyle\int_{-1}^{1}F(x)\,dx},
	\label{eq.dimless_gamma}
       \end{equation}
       to obtain $\lambda$ as a function of $\gamma$.
 \item Calculate other physical quantities from the expressions
       of~\eqref{eq.normalization_GY}, \eqref{eq.variables},
       \eqref{eq.effective_fermi_energy} and thermodynamic
       identities~\cite{Lieb} in the following form; 
       \begin{align}
	D &= \frac{Q}{\pi}\int_{-1}^{1}F(x)\,dx,  \label{eq.dimless_D} \\ 
	\frac{E^{\mathrm{eff}}_{\mathrm{F}}}{L}
	&= \frac{Q^3}{\pi}\int_{-1}^{1}x^2F(x)\,dx,
	\label{eq.dimless_eff_energy} \\
	\mu &=D^2\left(
	3e-\gamma\frac{de}{d\gamma}
	\right),\label{eq.formula_mu}\\
	v&=2D\left(3e-2\gamma\frac{de}{d\gamma}
	+\frac{1}{2}\gamma^2\frac{d^2e}{d\gamma^2}\right)^{1/2},
	\label{eq.formula_vs}
       \end{align}
       where $e\equiv E_{\mathrm{F}}
       /(ND^2)$,
       $\mu$ and $v$ are the scaled energy of the ground state, the
       chemical potential and the sound velocity, respectively.    
\end{enumerate}

\subsection{Strong coupling case}\label{subsec.strong}
We consider the large $\lambda$ case. In this case, only a few $a_n$'s
are enough to have good approximations. Indeed, it can be shown that 
$a_n=O(1/\lambda^{2n+1})$ for $n\ge 1$. Therefore, we only keep $a_0$,
$a_1$, and set $a_n\equiv 0$ $(n\ge 2)$,  
\begin{equation}
 F(x)=a_0 +a_1x^2.\label{eq.strong_F} 
\end{equation}
Equations~\eqref{eq.ap} give
\begin{subequations}
 \begin{align}
  a_0+\frac{2}{\pi}\arctan{\frac{1}{\lambda}}\cdot
  a_0+\frac{2}{\pi\lambda}\left(\frac{1}{3}-\frac{1}{5\lambda^2}\right)a_1
  =2, \\
  a_0 -\frac{2}{\pi\lambda^3}a_0-\frac{2}{3\pi\lambda^3}a_1=0, 
 \end{align}
\end{subequations}
and consequently,
\begin{subequations}\label{eq.strong_ap}
 \begin{align}
  a_0 &= 2\left(1-\frac{2}{\pi\lambda}+\frac{4}{\pi^2\lambda^2}
  +\frac{2\pi^2-24}{3\pi^3\lambda^3}\right), \\
  a_1 &= \frac{4}{\pi\lambda^3}. 
 \end{align}
\end{subequations}
Using~\eqref{eq.strong_F} with~\eqref{eq.strong_ap}
in~\eqref{eq.dimless_gamma}, we obtain $\lambda$ as a function of
$\gamma$. The result is
\begin{equation}
 \lambda =-\frac{2(2\gamma+1)}{\pi}\left(1 -\frac{\pi^2}{6(2\gamma
				    +1)^3}\right). 
 \label{eq.strong_lambda}
\end{equation}
The expressions of $a_0$ and $a_1$ in terms of $\gamma$ are
\begin{subequations}\label{eq.strong_F2}
\begin{align}
 a_0&=\frac{2\gamma+1}{\gamma}\left(1-\frac{\pi^2\gamma}{6(2\gamma
				+1)^4}\right), \\
 a_1&=-\frac{\pi^2}{2(2\gamma +1)^3}.
\end{align}
\end{subequations}
The ground state energy is calculated
from~\eqref{eq.dimless_eff_energy}, 
\begin{equation}
 \frac{E^{\mathrm{eff}}_{\mathrm{F}}}{L}
  =D^3\frac{\pi^2}{12}\left(\frac{\gamma}{2\gamma +1}\right)^2
 \left(1+\frac{4\pi^2}{15(2\gamma +1)^3}\right). \label{eq.strong_E}
\end{equation}
The cut-off rapidity $Q$, the chemical potential $\mu$ and the sound
velocity $v$ are given by 
\begin{align}
 Q&=\frac{\pi D}{2}\frac{\gamma}{2\gamma+1} \left(1+\frac{\pi^2}
 {6(2\gamma +1)^3}\right),\label{eq.strong_Q}\\
 \mu^{\mathrm{eff}}&=D^2\frac{\pi^2}{12}\frac{6\gamma +1}{2\gamma
 +1}\left(\frac{\gamma}{2\gamma +1}\right)^2 
 \left(1+\frac{8\pi^2}{5(2\gamma
 +1)^2(6\gamma+ 1)}\right),\label{eq.strong_mu}\\
 v&=2\pi D\left(\frac{\gamma}{2\gamma
 +1}\right)^2\left(1+\frac{2\pi^2}{3(2\gamma
 +1)^3}\right),\label{eq.strong_vs} 
\end{align}
where we have introduced the effective chemical potential
$\mu^{\mathrm{eff}}$, $\mu^{\mathrm{eff}}\equiv\mu +(1/4)D^2\gamma^2$.

\subsection{Weak coupling case}\label{subsec.weak}
This is the small $\lambda$ case. The lowest order approximation
of~\eqref{eq.ap} is
\begin{subequations}
 \label{eq.lowest_ap}
\begin{align}
 2a_0&=2,\\
 2a_p&=0,\quad p\ge 1.
\end{align}
\end{subequations}
A derivation of~\eqref{eq.lowest_ap} from~\eqref{eq.ap} is the same as
shown before~\cite{MW02,TI}. Another derivation is given in
Appendix \ref{ap.another_small_lambda}, where we treat the 
integral in a more direct manner in the small $\lambda$
case. Equations~\eqref{eq.lowest_ap} simply gives 
\begin{equation}
 F(x)=1.\label{eq.small_lowest_F}
\end{equation} 
We list here the analysis based on~\eqref{eq.small_lowest_F}:
\begin{align}
 \lambda&= -\frac{2}{\pi}\gamma,& Q&=\frac{\pi D}{2}, &&\\
 \frac{E^{\mathrm{eff}}_{\mathrm{F}}}{L}&=D^3\frac{\pi^2}{12}, 
 & \mu^{\mathrm{eff}}&=D^2\frac{\pi^2}{4}, &
 v&=\pi D.
\end{align}
It should be noted that, in contrast to the solution of the Lieb-Liniger
integral equation~\eqref{eq.LL}, all the coefficients $a_n$ are finite at
$\lambda\to 0$~\cite{MW02}. This nice analytic property can be seen
from~\eqref{eq.dimless_GY}, since the kernel
$\lambda/\{\pi[\lambda^2+(x-y)^2)]\}$ becomes an expression of
$\delta(x-y)$ in the limit $\lambda\to 0$. 

The second order approximation of~\eqref{eq.ap} is
\begin{equation}
2a_p+\frac{2}{\pi}\lambda\sum_{n=0}^{\infty}\frac{2p+1}{2n-2p-1}a_n
  =2\delta_{p0},\quad p\ge 0.
  \label{eq.second_ap}
\end{equation}
Here, $\delta_{p0}$ is the Kronecker's delta. A derivation
of~\eqref{eq.second_ap} is also given in the previous
papers~\cite{MW02,TI} and in Appendix~\ref{ap.another_small_lambda}. 
We set  
\begin{equation}
 a_n=\delta_{n0}+a^{(1)}_n+o(\lambda), \quad n\ge 0,\label{eq.ap_corrections} 
\end{equation} 
where $\delta_{n0}$ stands for the solution of~\eqref{eq.lowest_ap} and 
$a^{(1)}_n$ denote the corrections of the $\lambda^1$ order. Substitution 
of~\eqref{eq.ap_corrections} into~\eqref{eq.second_ap} yields
\begin{equation}
 2a^{(1)}_p-\frac{2}{\pi}\lambda =0, \quad p\ge 0, 
\end{equation}
which leads to
\begin{align} 
 F(x)&=1+\frac{\lambda}{\pi}\sum_{n=1}^{\infty}x^{2n}+o(\lambda)\nonumber\\ 
 &=1+\frac{\lambda}{\pi}\frac{1}{1-x^2}+o(\lambda). \label{eq.small_F}
\end{align}
This agrees with the approximation shown in the Gaudin's
book~\cite{Gaudin}. 
From~\eqref{eq.dimless_gamma}, $F(x)$ is integrated to give  
\begin{equation}
 \gamma
  =-\pi\lambda\left(2+\frac{2b}{\pi}\lambda\right)^{-1}, 
  \label{eq.gamma_lambda}  
\end{equation}
where the divergent sum $b$ is defined by
\begin{equation}
 b=\sum_{n=0}^{\infty}\frac{1}{2n+1}.
\end{equation}
Equation~\eqref{eq.gamma_lambda} is solved to obtain $\gamma$ as a
function of $\lambda$:  
\begin{equation}
 \lambda =-\frac{2}{\pi}\gamma\left(1-\frac{2b}{\pi^2}\gamma\right).
  \label{eq.lambda_gamma}
\end{equation}
 
We proceed to calculate the physical
quantities. From~\eqref{eq.dimless_D} and~\eqref{eq.lambda_gamma}, we
have 
\begin{equation}
 Q=\frac{\pi D}{2}\left(1-\frac{2b}{\pi^2}\gamma\right). 
\label{eq.small_Q}
\end{equation}
The ground state energy is obtained from~\eqref{eq.dimless_eff_energy},
\eqref{eq.small_F} and~\eqref{eq.small_Q}:
\begin{align}
 \frac{E^{\mathrm{eff}}_{\mathrm{F}}}{L}&=\frac{Q^3}{\pi}\left(\frac{2}{3}+ 
 \frac{2\lambda}{\pi}\sum_{n=0}^{\infty}\frac{1}{2n+3}\right)\nonumber\\
 &=D^3\frac{\pi^2}{12}\left(1+\frac{12}{\pi^2}\gamma\sum_{n=0}^{\infty}
 \frac{1}{(2n+1)(2n+3)}\right)\nonumber\\
 &=D^3\frac{\pi^2}{12}\left(1+\frac{6}{\pi^2}\gamma\right).
\end{align}
This recovers the previous result~\cite{KO} where the integrals are
directly treated without obtaining the explicit form of
$F(x)$. It is remarkable that we observe the cancellation of the
divergent sums as before~\cite{MW02, TI}. That is, the physical quantities
are expressed in power series of $\gamma$ without any divergent sum.   
The chemical potential and the sound velocity can be calculated
from the formulae~\eqref{eq.formula_mu} and~\eqref{eq.formula_vs},  
\begin{align}
 \mu^{\mathrm{eff}}&=D^2\frac{\pi^2}{4}\left(1+\frac{4}{\pi^2}\gamma
 \right),\\  
 v&=\pi D\left(1+\frac{\gamma}{\pi^2}\right). 
\end{align}
The limit $\gamma\to -0$ gives the well-known results for free fermions. 
\begin{figure}[htbp]
  \begin{center}
   \includegraphics[width=.85\linewidth]{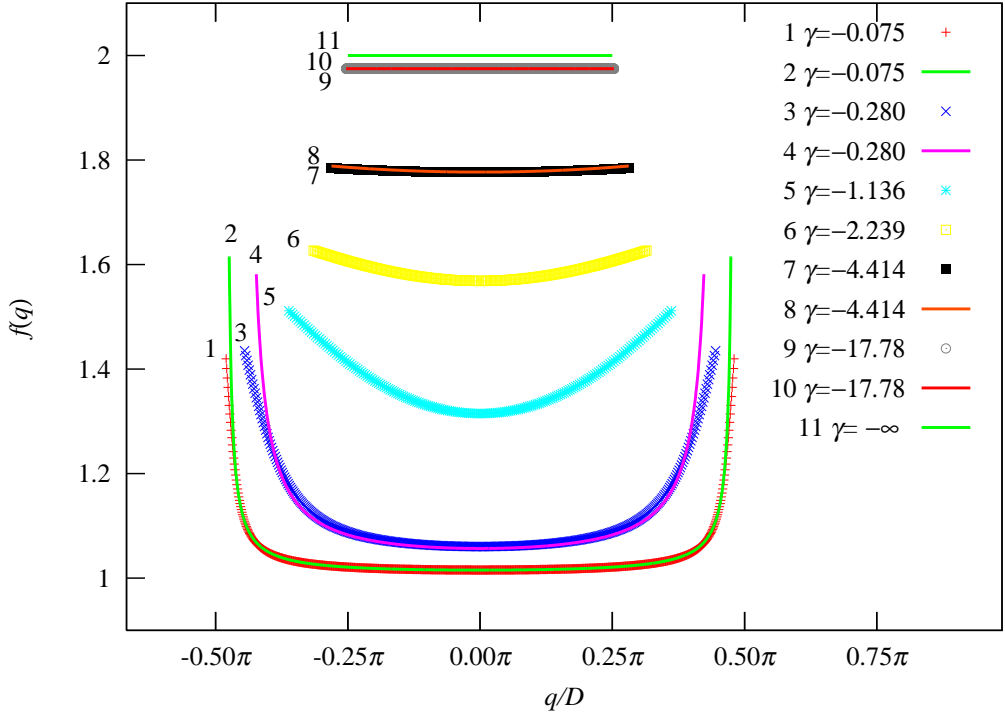}
   \caption{The normalized distribution function of spins $f(q)$
   in the ground state for various values of $\gamma$. Numerical
   results are indicated by points (\textcolor{red}{$+$},
   \textcolor{blue}{$\times$} etc.), 
   while the analytical ones, \eqref{eq.strong_F}
   and~\eqref{eq.small_F}, are by lines (\textbf{{\color{green}-----}, 
   {\color{red}-----}} etc). 
   Numerical solutions are obtained by a classical
   method~\cite{LL,KythePuri}. For small $\gamma$, analytical solutions
   are shown for the intervals $[-Q/D+|\gamma|/4,Q/D-|\gamma|/4]$ of
   $q/D$, which 
   correspond to $[-1+\lambda/4,1-\lambda/4]$ of $x$. For intermediate
   $\gamma$, $\gamma=-1.136$ and $\gamma=-2.239$, only the numerical
   results are shown. For large $\gamma$, analytical and numerical
   results are indistinguishable.}  
   \label{fig.fs}
  \end{center}
 \end{figure}%

In fig.~\ref{fig.fs}, we plot the
solutions~\eqref{eq.strong_F} with~\eqref{eq.strong_F2}
and~\eqref{eq.small_F} with~\eqref{eq.lambda_gamma}, for several values
of $\gamma$. For comparison, we also plot the numerical 
solutions based on the classical quadrature
method~\cite{LL,KythePuri}. In this figure, we choose a unit, 
$D=1$ so that $Q/D$ takes a value between $\pi/4$ and $\pi/2$ (see
~\eqref{eq.strong_Q} and~\eqref{eq.small_Q}). For a fixed $Q$ and $-1\le
x\le 1$, $q$ and $f(q)$ are determined by~\eqref{eq.variables} 
and~\eqref{eq.def_of_F}, respectively. Numerical values of $\gamma$ are
obtained through the relation $\gamma=-\lambda Q/D$. In the weak
coupling case, it is fair to say that the error is 
large near the end-points, $|q|/D\lessapprox
 Q/D$. This can be observed
analytically when the solution~\eqref{eq.small_F} is substituted into
the integral equation~\eqref{eq.dimless_GY}.

\section{The BCS-BEC Crossover}\label{sec.BCS-BEC}
Recently, there is a renewed interest in the crossover from a
Bardeen-Cooper-Schrieffer (BCS) superfluid with Cooper pairs to a
Bose-Einstein Condensate (BEC) of molecules composed of tightly bound
fermion pairs~\cite{ZW, Tokatly}. The experimental realizations of
low-dimensional trapped bose gases at ultracold 
temperatures also inspired the controversy even in
one-dimension~\cite{Batchelor, Fuchs}.    
The analytic solutions in \S~\ref{sec.solutions} of the Gaudin integral
equation and in the previous paper~\cite{MW02} of the Lieb-Liniger (LL)
integral equation enable us to make the quantitative discussion. To
avoid confusion, physical quantities, such as the mass $m$, the number
of particles $N$, the number density $D$, the coupling constants
$\lambda$, $\gamma$, and the ground state energy $E$ are denoted with
the subscripts $\mathrm{F}$ and $\mathrm{B}$ for a fermi gas with
spin-$1/2$ and a bose gas, respectively.   
       
The dimensionless form of the LL integral equation~\eqref{eq.LL}
reads~\cite{LL, MW02}  
\begin{gather}
 g(y)-\frac{\lambda_{\mathrm{B}}}{\pi}\int_{-1}^{1}\frac{g(x)\,dx}
 {\lambda_{\mathrm{B}}^2+(x-y)^2} 
  =\frac{1}{2\pi},  \label{eq.dimless_LL}\\
\intertext{where}
 g(x)=\rho(Kx),\qquad k=Kx, \qquad 2m_{\mathrm{B}}c=K\lambda_{\mathrm{B}}. 
\end{gather}
The relation between $\lambda_{\mathrm{B}}$ and $\gamma_{\mathrm{B}}$ is
\begin{equation}
 \gamma_{\mathrm{B}}=\frac{2m_{\mathrm{B}}c}{D_{\mathrm{B}}}=
  \frac{\lambda_{\mathrm{B}}}{\displaystyle\int_{-1}^{1}g(x)\,dx},
\end{equation}
where $D_{\mathrm{B}}=N_\mathrm{B}/L$. The ground state energy is given
by 
\begin{equation}
\frac{E_\mathrm{B}}{L}=\frac{K^3}{2m_{\mathrm{B}}}\int_{-1}^{1}x^2g(x)\,dx.  
\end{equation}
As already mentioned in the Gaudin's paper~\cite{Gaudin67},
strong coupling fermions described by the Gaudin integral
equation~\eqref{eq.dimless_GY} may be regarded as the repulsive bosons 
of dimers. To examine whether it is indeed true, we establish a 
relation between the exact solutions of the Gaudin integral 
equation~\eqref{eq.dimless_GY} and the LL integral
equation~\eqref{eq.dimless_LL}. In the following, we concentrate 
ourselves on the strong coupling case, i.e. the large
$\lambda_{\mathrm{F}}$ case for fermions and large
$\lambda_{\mathrm{B}}$ case for bosons are 
considered. For explicit comparison, we summarize here the results shown
in~\S~\ref{subsec.strong} and the previous papers~\cite{MW02, TI}. \\     
For $-\gamma_{\mathrm{F}}\gg 1$ (fermions with strong attraction), 
\begin{subequations}\label{eq.strong_fermi}
 \begin{align}
  f(q)&=\frac{2\gamma_{\mathrm{F}}
  +1}{\gamma_{\mathrm{F}}}\left( 
  1-\frac{\pi^2\gamma_{\mathrm{F}}}{6(2\gamma_{\mathrm{F}}
  +1)^4}\right) 
 -\frac{1}{\gamma_{\mathrm{F}}^3D_{\mathrm{F}}^2}q^2, \\
  \lambda_{\mathrm{F}} &=-\frac{4\gamma_{\mathrm{F}} +2}{\pi}\left(
  1-\frac{\pi^2}{6(2\gamma_{\mathrm{F}}
  +1)^3}\right),\\
  Q&=\frac{\pi
  D_\mathrm{F}}{2}\frac{\gamma_\mathrm{F}}{2\gamma_{\mathrm{F}}+1}
  \left(1+\frac{\pi^2}{6(2\gamma_{\mathrm{F}} 
  +1)^3}\right),\\
  \frac{E_\mathrm{F}^{\mathrm{eff}}}{L}&=
  \frac{D_{\mathrm{F}}^3}{2m_{\mathrm{F}}} 
  \frac{\pi^2}{12} 
  \left(\frac{\gamma_{\mathrm{F}}}{2\gamma_{\mathrm{F}}
  +1}\right)^2 
 \left(1+\frac{4\pi^2}{15(2\gamma_{\mathrm{F}}
  +1)^3}\right),\label{eq.E_fermions}\\
  \mu_{\mathrm{F}}^{\mathrm{eff}}&=\frac{D_{\mathrm{F}}^2}{2m_{\mathrm{F}}}
  \frac{\pi^2}{12}\frac{6\gamma_{\mathrm{F}}  +1}{2\gamma_{\mathrm{F}} 
  +1}\left(\frac{\gamma_{\mathrm{F}}}{2\gamma_{\mathrm{F}} +1}\right)^2 
  \left(1+\frac{8\pi^2}{5(2\gamma_{\mathrm{F}}
  +1)^2(6\gamma_{\mathrm{F}}+ 1)}\right),\label{eq.mu_fermions}\\
  v_{\mathrm{F}}&=\frac{2\pi
  D_{\mathrm{F}}}{2m_{\mathrm{F}}}\left(
  \frac{\gamma_{\mathrm{F}}}{2\gamma_{\mathrm{F}}  
  +1}\right)^2\left(1+\frac{2\pi^2}{3(2\gamma_{\mathrm{F}}
  +1)^3}\right).\label{eq.v_fermions}
 \end{align} 
\end{subequations}
 For $\gamma_{\mathrm{B}}\gg 1$ (bosons with strong repulsion), 
\begin{subequations}\label{eq.strong_bose}
\begin{align}
 \rho(k)&=\frac{\gamma_{\mathrm{B}}
 +2}{2\pi\gamma_{\mathrm{B}}} 
 \left(1-\frac{2\pi^2}{3\gamma_{\mathrm{B}}
 (\gamma_{\mathrm{B}}+2)^2}\right)   
 -\frac{1}{\pi\gamma^3_{\mathrm{B}}D^2_{\mathrm{B}}}k^2,\\ 
 \lambda_{\mathrm{B}} &=\frac{\gamma_{\mathrm{B}} +2}{\pi}\left(1
 -\frac{4\pi^2}{3(\gamma_{\mathrm{B}} 
 +2)^3}
 \right), \\
 K&=\pi
 D_\mathrm{B}
 \frac{\gamma_{\mathrm{B}}}{\gamma_{\mathrm{B}}+2}\left(1+\frac{4\pi^2}
 {3(\gamma_\mathrm{B}+2)^3}\right),\\
 \frac{E_\mathrm{B}}{L}&=\frac{D_{\mathrm{B}}^3}{2m_{\mathrm{B}}}
 \frac{\pi^2}{3}\left(    
 \frac{\gamma_{\mathrm{B}}}{\gamma_{\mathrm{B}}
 +2}\right)^2 
 \left(1+\frac{32\pi^2}{15(\gamma_{\mathrm{B}}
 +2)^3}\right),\label{eq.E_bosons}\\
 \mu_{\mathrm{B}}&=\frac{D_{\mathrm{B}}^2}{2m_{\mathrm{B}}}
 \frac{3\gamma_{\mathrm{B}}+2}{\gamma_{\mathrm{B}}+2}\frac{\pi^2}{3}
 \left(\frac{\gamma_{\mathrm{B}}}{\gamma_{\mathrm{B}}+2}\right)^2\left(
 1+\frac{64\pi^2}{5(\gamma_{\mathrm{B}}+2)^2(3\gamma_{\mathrm{B}}+2)}\right), 
 \label{eq.mu_bosons}
 \\
 v_{\mathrm{B}}&=\frac{2\pi D_{\mathrm{B}}}{2m_{\mathrm{B}}}\left(
 \frac{\gamma_{\mathrm{B}}}{\gamma_{\mathrm{B}}+2}\right)^2
 \left(1+\frac{16\pi^2}{3(\gamma_{\mathrm{B}}+2)^3}\right).
 \label{eq.v_bosons}
\end{align}
\end{subequations}
Identifying $N_{\mathrm{F}}$ fermions with mass
$m_{\mathrm{F}}$ with $N_{\mathrm{B}}$ bosons of dimers with mass
$m_{\mathrm{B}}$, we have    
\begin{subequations}\label{eq.fermi_bose}
 \begin{align}
  m_{\mathrm{B}}&=2m_{\mathrm{F}}, &
  N_{\mathrm{B}}&=\frac{N_{\mathrm{F}}}{2}, &
  D_{\mathrm{B}}&=\frac{D_{\mathrm{F}}}{2}. \\
\intertext{Hence, from~\eqref{eq.def_gamma} we have} 
  && \gamma_{\mathrm{B}}&
  \lefteqn{=4\gamma_{\mathrm{F}}.} &&
 \end{align} 
\end{subequations}
Substitution of~\eqref{eq.fermi_bose} into~\eqref{eq.E_fermions} gives
\begin{equation}
 \begin{split}
  \frac{E_{\mathrm{F}}^{\mathrm{eff}}}{L}&=
  \frac{(2D_{\mathrm{B}})^3}{2m_{\mathrm{B}}/2}
  \frac{\pi^2}{12}\left(
  \frac{\gamma_{\mathrm{B}}/4}{\gamma_{\mathrm{B}}/2
  +1}\right)^2
  \left(1+\frac{4\pi^2}{15(\gamma_{\mathrm{B}}/2
  +1)^3}\right) \\
  &=\frac{D_{\mathrm{B}}^3}{2m_{\mathrm{B}}}
  \frac{\pi^2}{3}\left(\frac{\gamma_{\mathrm{B}}}
  {\gamma_{\mathrm{B}}+2}\right)^2
  \left(1+\frac{32\pi^2}{15(\gamma_{\mathrm{B}}+2)^3}\right) 
  =\frac{E_{\mathrm{B}}}{L}. 
 \end{split}
\end{equation}
The effective ground state energies are exactly the same through the
relation~\eqref{eq.fermi_bose} in the strong coupling case. 
We can easily verify that other physical quantities are also smoothly   
connected through the relation~\eqref{eq.fermi_bose} at
the point $1/\gamma_{\mathrm{F}}=1/\gamma_{\mathrm{B}}=0$. 
In fact, in the case being considered, equations~\eqref{eq.dimless_GY} 
and~\eqref{eq.dimless_LL} can be treated in a unified way, if we regard
$\lambda$ as a parameter and $D$ as a normalization condition for these
integral equations. We perform such calculations in
Appendix~\ref{app.infinity}. 

We note that the quasi-momenta for bosons and fermions are related as 
\begin{align}
 K=2Q,\qquad k=2q. \label{eq.bound_q}
\end{align}
These relations are consistent with the assumptions that lead us to
the Gaudin integral equation. For negative $c$, quasi-momentum
$\{k\}$ of the fermions (not to be confused with $k$ for the repulsive
bosons in~\eqref{eq.bound_q}) become imaginary. In the thermodynamic 
limit, two $k$ s with a real part $q$, which are complex conjugates of
each other, form a bound state of two fermions with the binding energy
$c^2/2$ (see~\eqref{eq.fermi_energy})~\cite{Takahashi}.  This kind of 
argument usually starts with the $c= -0$ limit. Remarkably, it holds
also in the strong coupling case.  

Equations~\eqref{eq.E_fermions} and~\eqref{eq.E_bosons} need some
explanation. Since we are seeking for the solution of $O(1/\lambda^3)$,
the expansion must be in powers of either $1/\lambda\simeq
1/(\gamma_{\mathrm{B}}+2)\simeq1/(\gamma_{\mathrm{F}}+1/2)$ or in
$1/\gamma_{\mathrm{B}}\simeq 1/\gamma_{\mathrm{F}}$. However, it occurs
that the solution of the first term should be
$(\gamma_{\mathrm{B}}/(\gamma_{\mathrm{B}}+2))^2
\simeq(|\gamma_{\mathrm{F}}|/(|\gamma_\mathrm{F}|-1/2))^2$, not be
simply $1$. This factor can be counted as the reminiscent of the
effective ``diameter'' $a$ of each particle, in the limit of the 
hard-core bose gas, or, in other words, the quantized version of the
classical Tonks gas~\cite{Tonks}. In fact, the energy at zero
temperature for the hard-core bose gas is calculated as~\cite{WK02} 
\begin{equation}
 \frac{E_\mathrm{hc}}{L}=\frac{D_{\mathrm{B}}^3}{2m_{\mathrm{B}}}
  \frac{\pi^2}{3}\left(\frac{1}{1-aD_{\mathrm{B}}}\right)^2.
  \label{eq.E_hc}   
\end{equation}
Besides, equating the scattering phase shifts in the low
energy limit, we find~\cite{MW03}
\begin{equation}
 a=-\frac{1}{2m_{\mathrm{B}}}\frac{2}{c}=-\frac{1}{2m_{\mathrm{F}}}
  \frac{1}{c}.
\label{eq.diameter}
\end{equation} 
With the relation~\eqref{eq.diameter}, we see that~\eqref{eq.E_hc} and
the first terms of~\eqref{eq.E_bosons} and~\eqref{eq.E_fermions} become
exactly the same.  
The positiveness (negativeness) of the diameter $a$ for the attractive
fermions (repulsive bosons) is due to its statistical property. 
The negativeness for the repulsive bosons can also be interpreted as the 
transparency effect for a finite $c>0$. It is easy to verify that the
chemical potential $\mu_\mathrm{hc}$ and the sound velocity
$v_\mathrm{hc}$ of the hard-core bose gas~\cite{WK02} are obtained by
the substitution of~\eqref{eq.diameter} with~\eqref{eq.fermi_bose} into 
either~\eqref{eq.mu_fermions} and~\eqref{eq.v_fermions}
or~\eqref{eq.mu_bosons} and~\eqref{eq.v_bosons}. Discarding the second
term in the last brackets, we obtain
\begin{align}
 \mu_\mathrm{hc} &
 =\frac{D_{\mathrm{B}}^2}{2m_\mathrm{B}}\frac{\pi^2}{3} 
 \frac{3-aD_{\mathrm{B}}}{1-aD_{\mathrm{B}}} 
 \left(\frac{1}{1-aD_{\mathrm{B}}}\right)^2, \\
 v_{\mathrm{hc}} &=\frac{2\pi D_{\mathrm{B}}}{2m_{\mathrm{B}}}
 \left(\frac{1}{1-aD_{\mathrm{B}}}\right)^2.
\end{align} 
Note that $\mu_{\mathrm{hc}}$ corresponds to
$2\mu_{\mathrm{F}}^{\mathrm{eff}}$. 

Now we are ready to describe the BCS-BEC crossover in one-dimension. The
discussion above for smooth connections in the strong 
coupling cases justifies a link between the two systems. In view of the
physical quantities, we can map two fermions of 
opposite spins forming a bound state whose quasi-momenta are $q\pm ic/2$
onto one boson with quasi-momentum $k=2q$. In such a way, the effective
ground state energy per volume, the chemical potential and the velocity
of sound can be identified.    
If we take $\gamma_{\mathrm{F}}$ as the coupling constant common to the
both systems and vary $1/\gamma_{\mathrm{F}}$ from $-\infty$ to
$+\infty$ through $0$, we find the following asymptotic expressions
of the effective energy:  
\begin{align}
  \frac{E_{\mathrm{F}}^{\mathrm{eff}}}{N\epsilon_\mathrm{FF}}
 &=
 \begin{cases}
   \displaystyle \frac{1}{3}\left(1+ \frac{6}{\pi^2}
  \gamma_{\mathrm{F}}\right), & 
   -1/\gamma_{\mathrm{F}} \gg 1, \\
   \displaystyle \frac{1}{3}
  \left(\frac{\gamma_{\mathrm{F}}}{2\gamma_{\mathrm{F}} +1}\right)^2
  \left(1+ \frac{4\pi^2}{15(2\gamma_{\mathrm{F}} +1)^3}\right),
  &  1/\gamma_{\mathrm{F}} \sim 0,\\
   \displaystyle \frac{\gamma_{\mathrm{F}}}{\pi^2}
  \left(1 - \frac{8}{3\pi}\gamma_{\mathrm{F}}^
  {1/2}\right), &
   1/\gamma_{\mathrm{F}} \gg 1,
 \end{cases}
\end{align}
where $\epsilon_\mathrm{FF}$ is the Fermi energy of the free fermions, 
\begin{equation}
   \epsilon_\mathrm{FF}\equiv \frac{1}{2m_\mathrm{F}}
  \left( \frac{\pi D_\mathrm{F}}{2}\right)^2.
\end{equation}

\begin{figure}[htbp]
  \begin{center}
   \includegraphics[width=0.82\linewidth]{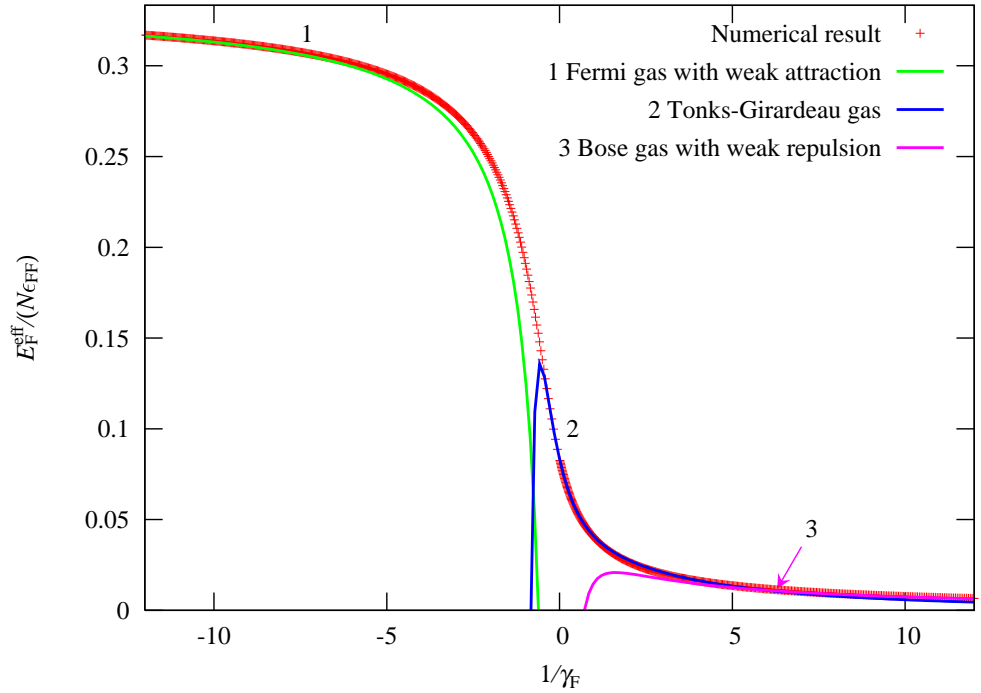}
   \caption{Asymptotics of the ground state energy
   $E_{\mathrm{F}}^{\mathrm{eff}}$ in units of the Fermi energy of the
   free fermions as a functions of the inverse of the coupling constant,
   $1/\gamma_{\mathrm{F}}$. }   
   \label{fig.energy}
  \end{center}
 \end{figure}
The meanings of the above three expressions are clear. They are the
energy of the Hartree-Fock approximation for weakly interacting 
fermions~\cite{KO}, the energy of either the hard-core bose gas or the
spinless non-interacting fermions~\cite{Girardeau}, and the energy due
to the Bogoliubov theory for weakly interacting bosons~\cite{LL}, for
$-1/\gamma_{\mathrm{F}}\gg 1$, 
$1/\gamma_{\mathrm{F}}\sim 0$, and $1/\gamma_{\mathrm{F}}\gg 1$,
respectively. The asymptotic 
solutions of the LL and the Gaudin integral equation reproduce the
results of these three ``phases'' rigorously. Two crossovers
for the intermediate value of $\gamma_{\mathrm{F}}$ are by no means a phase
transition. The solutions are continuous and analytic functions of
$1/\gamma_{\mathrm{F}}$ except the two points $-\infty$ and
$+\infty$. Those can be proved mathematically from the properties of the
integral equations.  

\begin{center}
\begin{figure}[htbp]
\begin{center}
 {\scriptsize
 \unitlength 0.1in
 \begin{picture}( 48.8200, 14.1500)(  1.6500,-24.1500)
  \special{pn 20}%
  \special{pa 568 2200}%
  \special{pa 5048 2200}%
  \special{fp}%
  \put(44.0700,-18.0000){\makebox(0,0){Bose Gas}}%
  \put(44.0700,-15.0000){\makebox(0,0){$\delta$-Function}}%
  \put(10.4700,-18.0000){\makebox(0,0){Fermi Gas}}%
  \put(10.4700,-15.0000){\makebox(0,0){$\delta$-Function}}%
  \put(27.2700,-16.0000){\makebox(0,0){(Hard-Core Bose Gas)}}%
  \put(27.1000,-25.0000){\makebox(0,0){$1/\gamma_{\mathrm{F}}$}}%
  \put(10.4700,-12.0000){\makebox(0,0){Weak Attractive}}%
  \put(44.0700,-12.0000){\makebox(0,0){Weak Replusive}}%
  \put(27.2700,-13.0000){\makebox(0,0){Tonks-Girardeau Gas}}%
  \put(27.0000,-23.1000){\makebox(0,0){0}}%
  \special{pn 8}%
  \special{pa 3768 2200}%
  \special{pa 3768 1000}%
  \special{dt 0.045}%
  \special{pn 8}%
  \special{pa 1688 2200}%
  \special{pa 1688 1000}%
  \special{dt 0.045}%
  \special{pn 20}%
  \special{pa 4088 2200}%
  \special{pa 4088 2200}%
  \special{fp}%
  \special{pa 4088 2200}%
  \special{pa 4088 2200}%
  \special{fp}%
  \put(50.6000,-23.0000){\makebox(0,0){$+\infty$}}%
  \put(5.7000,-23.1000){\makebox(0,0){$-\infty$}}%
  \special{pn 20}%
  \special{pa 2700 2192}%
  \special{pa 2700 2072}%
  \special{fp}%
 \end{picture}%
 ~\\
 }
 \caption{Three regimes of one-dimensional quantum gas: for
 $\gamma_\mathrm{F}<0$, spin-$1/2$ fermi gas and for
 $\gamma_\mathrm{F}=\gamma_\mathrm{B}/4>0$, $\delta$-function bose gas.}
 \label{fig.phase}
\end{center}
\end{figure}
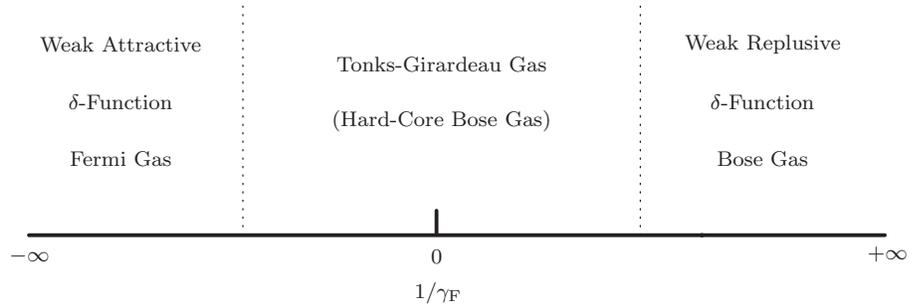%
\end{center}

Let us explain these three regimes in detail. 
When $-1/\gamma_{\mathrm{F}}\gg 1$, the ground state is a BCS-like state
with Cooper pairs~\cite{KO}. The effective size of this bound state is
much larger than the average interparticle distance. The existence 
of the thermodynamic limit is guaranteed by the Pauli exclusion principle;
i.e. there is no bound state of more than two fermions.  

The intermediate regime of the hard-core bose gas around
$1/\gamma_{\mathrm{F}} =0$ should be 
distinguished from the Girardeau gas, which corresponds only to the
limit $1/\gamma_{\mathrm{F}} =0$.  Indeed, when we compare our
result with the numerical one, as in fig.~\ref{fig.energy}, it traces
well the sudden change around $1/\gamma_{\mathrm{F}}=0$. 
In this region of the bosonic side, the strong mutual repulsion compels 
the bosonic particles to exhibit the fermionic properties due to
one-dimensional kinematics. On the other hand, even the infinitely
strong attraction cannot causes the fermions to condense. Namely,
the distribution and the energy spectrum of the tightly
bound fermion pairs can be identified with those of spinless hard-core
bosons with a positive ``diameter''. 
    
In the regime of weakly interacting bose gas, the lowest order solution
of the distribution $\rho(k)$ is~\cite{LL,MW02}
\begin{equation} 
 \rho(k)=\frac{1}{2\pi D_{\mathrm{F}}\gamma_{\mathrm{F}}}
  \left(D_\mathrm{F}^2\gamma_{\mathrm{F}} -q^2\right)^{1/2}. 
\end{equation}
This indicates that the weaker the interaction, the narrower the
quasi-momentum distribution region and the higher the peak of the 
distribution. In the recent experiment this behavior was observed for
$\gamma_\mathrm{B}= 0.5$~\cite{Pardes}. In the limit 
$1/\gamma_{\mathrm{F}}\to +\infty$, all the particles condense into the
zero quasi-momentum state. 

In the same manner as we have done in the effective ground state energy,
we list here the asymptotic expressions of the velocity of sound and the 
effective chemical potential:
\begin{align} 
 \frac{v_\mathrm{F}}{v_{\mathrm{FF}}}&
 =
 \begin{cases}
  \displaystyle 1+\frac{\gamma_{\mathrm{F}}}{\pi^2}, 
  & -1/\gamma_{\mathrm{F}}\gg 1, \\
  \displaystyle 2\left(\frac{\gamma_{\mathrm{F}}}{2\gamma_{\mathrm{F}}
  +1}
  \right)^2
  \left(1+\frac{2\pi^2}{3(2\gamma_{\mathrm{F}} +1)^3}\right), &
  1/\gamma_{\mathrm{F}}\sim 0, \\
  \displaystyle \frac{1}{\pi^2}\gamma_{\mathrm{F}}^{1/2}\left(1-\frac{1}{2\pi}
  \gamma_{\mathrm{F}}^{1/2}\right), & 1/\gamma_{\mathrm{F}}\gg 1.
 \end{cases}
\end{align}
where $v_{\mathrm{FF}}$ is the Fermi velocity of the free fermions,
\begin{align}
 v_{\mathrm{FF}}&\equiv\frac{\pi D_\mathrm{F}}{2m_\mathrm{F}}.
\end{align}
\begin{figure}[htbp]
  \begin{center}
   \includegraphics[width=0.82\linewidth]{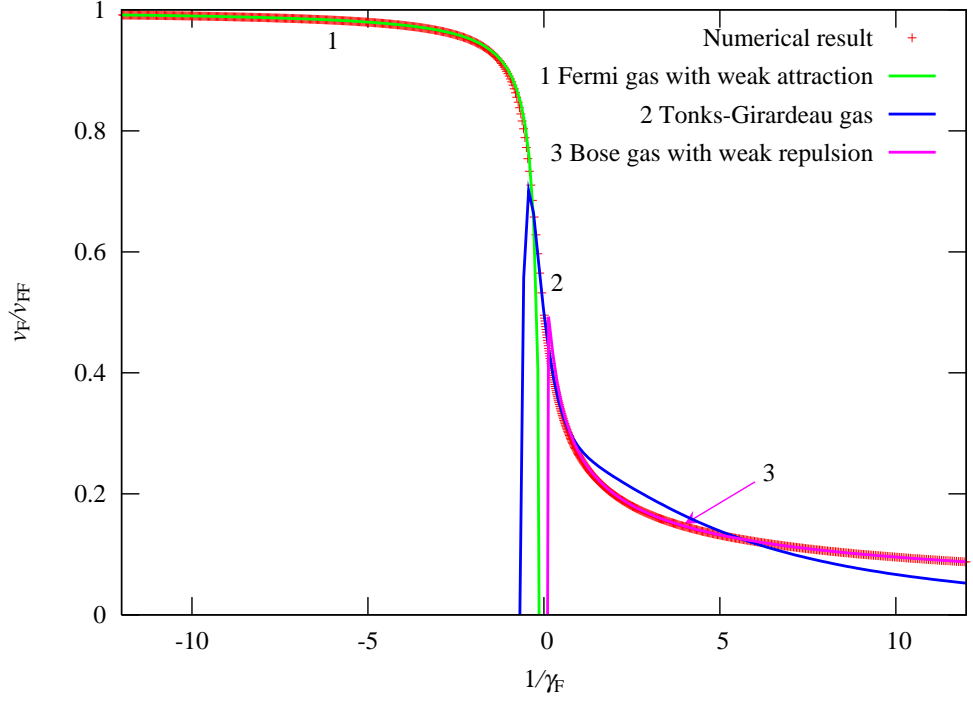}
   \caption{Asymptotics of the sound velocity $v_\mathrm{F}$ in units of 
   the Fermi velocity of the free fermions $v_{\mathrm{FF}}$ as a 
   function of the inverse of the coupling constant,
   $1/\gamma_{\mathrm{F}}$. }   
   \label{fig.v}
  \end{center}
 \end{figure}%
and
\begin{align}
 \frac{\mu^{\mathrm{eff}}_\mathrm{F}}{\epsilon_\mathrm{FF}}&=
 \begin{cases}
 \displaystyle  1+\frac{4}{\pi^2}\gamma_{\mathrm{F}}, 
  & -1/\gamma_{\mathrm{F}}\gg 1,\\
  \displaystyle  \frac{1}{3}
  \frac{6\gamma_{\mathrm{F}} +1}{2\gamma_{\mathrm{F}} +1}\left(
  \frac{\gamma_{\mathrm{F}}}{2\gamma_{\mathrm{F}} +1}
  \right)^2\left(1+\frac{8\pi^2}
  {5(2\gamma_{\mathrm{F}} +1)^2(6\gamma_{\mathrm{F}} +1)}\right), &
  1/\gamma_{\mathrm{F}}\sim 0, \\
  \displaystyle
  \frac{2}{\pi^2}\gamma_{\mathrm{F}}\left(1-\frac{2}{\pi}
  \gamma_{\mathrm{F}}^{1/2}\right), 
  & 1/\gamma_{\mathrm{F}}\gg 1.
 \end{cases}
\end{align}
\begin{figure}[htbp]
  \begin{center}
   \includegraphics[width=0.82\linewidth]{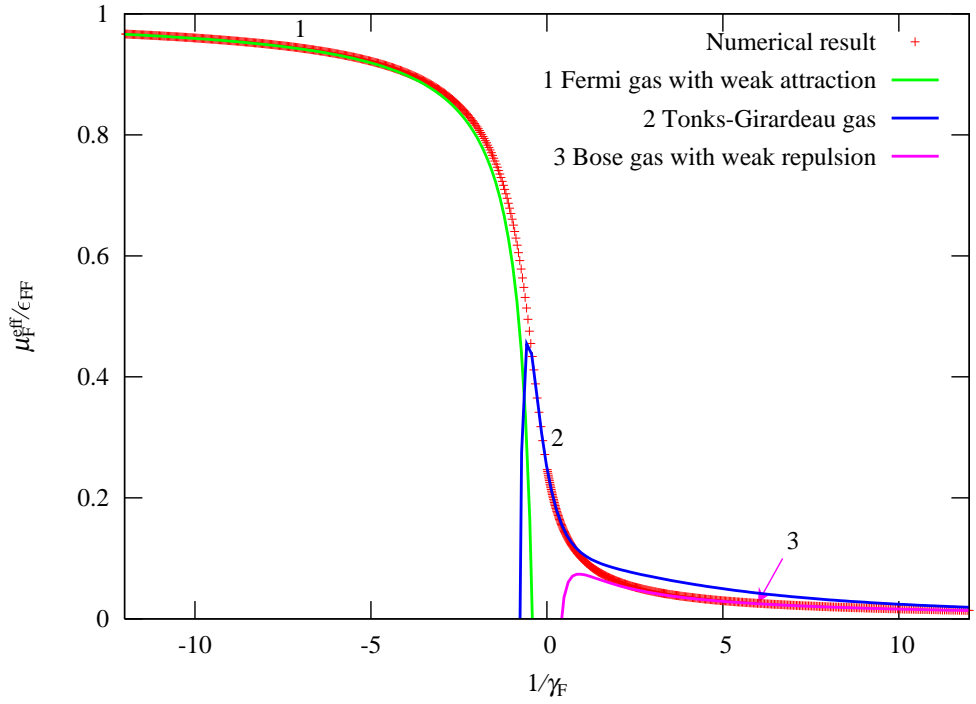}
   \caption{Asymptotics of the chemical potential 
   $\mu_{\mathrm{F}}^{\mathrm{eff}}$ in units of the Fermi energy of the
   free fermions $\epsilon_{\mathrm{FF}}$ as a function of the inverse
   of the coupling constant, $1/\gamma_{\mathrm{F}}$. }  
   \label{fig.mu}
  \end{center}
 \end{figure}%
These expressions are plotted in
fig.~\ref{fig.v} and fig.~\ref{fig.mu} as functions of
$1/\gamma_{\mathrm{F}}$.

\section{Analogy with the Circular Plate Condenser}\label{sec.condenser}
An intriguing fact is that the Gaudin integral equation and the LL
integral equation also appear in a completely different subject, the 
circular plate condenser~\cite{Love, Gaudin71, Gaudin}. We consider the
condenser of the radius $1$ whose two plates are 
separated by a distance $\lambda$.   
Cylindrical coordinates are taken as follows: the radial distance from
the common axis of the disks is taken to be $\rho$, the distances from
the disks to be $\zeta$, $\zeta'$, respectively. 
E. R. Love~\cite{Love} (1949) proved that the potential 
$\phi(\rho, \zeta, \zeta')$ due to the disks is expressed as
\begin{equation}
 \phi(\rho,\zeta,\zeta')=\frac{V_0}{\pi}\int_{-1}^{1}
  \left\{\frac{1}{\left[\rho^2+(\zeta+it)^2\right]^{1/2}}\pm\frac{1}
   {\left[\rho^2+(\zeta'-it)^2 \right]^{1/2}}\right\}h(t)\,dt,
  \label{eq.potential}
\end{equation} 
where $h(t)$ is determined by
\begin{equation}
 h(t)\pm\frac{\lambda}{\pi}\int_{-1}^{1}\frac{h(s)\,ds}{(t-s)^2+\lambda^2}
  =1.\label{eq.Love}
\end{equation}
The capacitance $C$ of the condenser is given by 
\begin{equation}
 C=\frac{1}{2\pi}\int_{-1}^{1}h(t)\,dt.\label{eq.capacitance}
\end{equation}
In the above expressions, the upper signs refer to the case of equally
charged disks at potential $V_0$, and the lower signs to that of
oppositely charged disks, namely, disks at potential $V_0$ and $-V_0$. 
The Love equation~\eqref{eq.Love} is exactly in the same form as the
Gaudin and the LL integral equation:  to be precise, $h(x)$
in~\eqref{eq.Love} corresponds to $(1/2)F(x)$ and $2\pi g(x)$
in~\eqref{eq.dimless_GY} and~\eqref{eq.dimless_LL}, respectively.   

In the case of oppositely
charged disks (lower signs), the solution of~\eqref{eq.Love} is calculated
for small $\lambda$ (small separation). By the heuristic
argument~\cite{Hutson, Popov}, away from the edges $h(t)$ to
the $o(1)$ order is written as
\begin{equation}
 h(t)=\frac{1}{\lambda}(1-t^2)^{1/2}+\frac{(1-t^2)^{-1/2}}
  {2\pi}\left(\log\frac{16\pi e}{\lambda}
	   -t\log\frac{1+t}{1-t}\right)+o(1),\label{hutson}
\end{equation}
and with~\eqref{eq.capacitance}, the capacitance is 
\begin{equation}
 C=\frac{1}{4\lambda}+\frac{1}{4\pi}\log\frac{16\pi}{\lambda e}+o(1).
\end{equation}
This result is originally by G. Kirchhoff (1877)~\cite{Kirchhoff}.

On the other hand, we have~\cite{MW02,MW03} 
\begin{align}
 g(x)=
 \frac{1}{2\pi\lambda}
 (1-x^2)^{1/2}
 -\frac{1}{4\pi}\sum_{n=0}^{\infty}\sum_{m=0}^{\infty}
	\frac{(2m+1)!!\,\,(2m-1)!!\,\,(2n-1)!!}{\left[(2m)!!\right]^2
	(2n)!!\,\,(2n-2m-1)}x^{2n},\label{ourrho}
\end{align}
and
\begin{align}
 C&=\int_{-1}^{1}g(x)\,dx
 =\frac{1}{4\lambda}-\frac{1}{2\pi}+\frac{1}{4}\mathfrak{b},
\quad \text{where}\quad 
 \mathfrak{b}=\sum_{m=0}^{\infty}
  \left[\frac{(2m-1)!!}{(2m)!!}\right]^2.
\end{align}
We regularize the divergent sum $\mathfrak{b}$ as follows. We set  
\begin{equation} 
 \mathfrak{b}(k)\equiv\sum_{m=0}^{\infty}
  \left[\frac{(2m-1)!!}{(2m)!!}\right]^2k^{2m} 
  =\frac{2}{\pi}K(k).\label{div_sum1}
\end{equation}
The function $K(k)$ is the complete elliptic integral of the first kind
with the modulus $k$.  Asymptotic behavior of $K(k)$ is
\begin{equation}
 K(k)=\frac{1}{2}\log\frac{16}{1-k^2}\quad \text{as} \quad
  k^2\nearrow 1.\label{asymelliptic}
\end{equation}
If we choose $1-k^2=\lambda/(\pi e)$ in~\eqref{asymelliptic}
with~\eqref{div_sum1}, we get   
\begin{equation}
\mathfrak{b}
  =\lim_{k^2\nearrow 1}\mathfrak{b}(k)
  =\frac{1}{\pi}\log\frac{16\pi e}{\lambda}. 
  \label{div_sum2}
\end{equation}
Using the expression~\eqref{div_sum2}, we may rewrite~\eqref{ourrho} as 
\begin{align}
 g(x)&=\frac{1}{2\pi\lambda}(1-x^2)^{1/2}+\frac{1}{4\pi}
 \sum_{l=0}^{\infty}\sum_{m=0}^{\infty}\left[\frac{(2m-1)!!}
 {(2m)!!}\right]^2\frac{(2l-1)!!}{(2l)!!}x^{2l}\nonumber\\
 &\quad +\frac{1}{4\pi}
 \sum_{l=0}^{\infty}\sum_{m=0}^{\infty}\frac{(2l-1)!!\,2l}{(2l)!!}
 \left[\frac{(2m-1)!!}{(2m)!!}\right]^2\frac{1}{2m-(2l-1)}x^{2l}\nonumber\\
 &=\frac{1}{2\pi\lambda}(1-x^2)^{1/2}
 +\frac{1}{4\pi^2}\left(\log\frac{16\pi e}{\lambda}
 -x\log\frac{1+x}{1-x}\right)\frac{1}{(1-x^2)^{1/2}}.\label{ourrho2}
\end{align}
Here, we have used the power series formulae, 
\begin{align}
 \sum_{p=0}^{\infty}\dfrac{(2p-1)!!}{(2p)!!}x^{2p}&=(1-x^2)^{-1/2}, \quad
 |x|<1,\label{eq.1-x_-1/2}
 \\
 \sum_{n=0}^{\infty}\dfrac{1}{2n+1}x^{2n+1}&=\dfrac{1}{2}\log\dfrac{1+x}{1-x},  
 \quad |x|<1,\label{eq.logx}
\end{align}
and the following summation formula, 
\begin{multline}
 \sum_{m=0}^{\infty}\left[\frac{(2m-1)!!}{(2m)!!}\right]^2
 \frac{1}{2m-(2l+1)}
 =-\frac{2}{\pi}\frac{(2l)!!}{(2l+1)!!}\sum_{k=0}^{l}
 \frac{1}{2l-2k+1}\frac{(2k-1)!!}{(2k)!!},\\
 l=0,1,2,\ldots,\label{newsum1}
\end{multline}
which is proved in Appendix~\ref{app.pf_sum}. 

Therefore, by the regularization of the divergent sum~\eqref{div_sum2},
we exactly reproduce Hutson's solution~\eqref{hutson} for $x\sim 0$
for $o(1)$. 
This coincidence is reasonable since we have started from the expansion
of $g(x)$ around $x=0$. 

We may also calculate the physical quantities  
from~\eqref{ourrho2}. At our consideration, to the $\lambda^0$
order, we can integrate~\eqref{ourrho2} for the whole interval
$[-1,1]$. Indeed, the corrections come from near the edges,
$x\in[-1,-1+\lambda]\cup[1-\lambda,1]$,  and $g(x)$ is estimated to be
of the order $1/\sqrt{\lambda}$ there. Then, the corrections are of the 
order $\sqrt{\lambda}$, and thus can be neglected.     

The case of equally charged disks (upper signs in~\eqref{eq.potential}
and~\eqref{eq.Love}) has not attracted much interests. So 
far, no prescription has been proposed to solve~\eqref{eq.Love}
asymptotically. Our solution~\eqref{eq.small_F} gives    
\begin{equation}
 C=\frac{1}{4\pi}\int_{-1}^{1}F(x)\,dx
  =\frac{1}{2\pi}+\frac{\lambda}{4\pi^2}\log\frac{4\pi
  e}{\lambda}+O(\lambda). \label{eq.eq_capacitance}
\end{equation}
To obtain the last expression, the regularization of the 
divergent sum $b$ in the same manner as~\eqref{div_sum2}
is required: 
\begin{equation}
 b=\sum_{n=0}^{\infty}\frac{1}{2m+1}=\lim_{k^2\nearrow 
  1}\sum_{m=0}^{\infty}\frac{1}{2m+1}k^{2m} 
  =\frac{1}{2}\log\frac{4\pi e}{\lambda}, 
\end{equation}
where~\eqref{eq.logx} has been used to obtain the last expression. 
The first term in~\eqref{eq.eq_capacitance} was also referred to
in the Kirchhoff's paper~\cite{Kirchhoff}. The second term agrees with
the analysis by Gaudin~\cite{Gaudin} except $4$ in the logarithmic
function.

\section{Concluding Remarks}
We have presented solutions of the Gaudin integral
equation~\eqref{eq.GY}. We have solved the integral equation in the 
form of power series.  
Based on the solution, the strong coupling case is analyzed, in
relation to the Lieb-Liniger integral equation for the bosons with
repulsive interaction. Then, we have a unified view on $\delta$-function
attractive fermions and repulsive bosons in one-dimension. Main results
are summarized as follows.   
\begin{enumerate}[i)]
 \item As to the thermodynamic quantities, the expressions in
       the strong coupling cases $|\gamma|\gg 1$ are identical through
       the relations~\eqref{eq.fermi_bose}. 
 \item The limit $1/\gamma=0$ yields the Girardeau gas~\cite{Girardeau},
       in the sense that the energy spectra coincide with that of
       free fermions.  
 \item Around the point $1/\gamma =0$, the system behaves like a
       hard-core bose gas with a diameter $a$ given
       by~\eqref{eq.diameter}. This indicates the existence of dimers of
       tightly bound fermions 
       ($1/\gamma= -0$, $a>0$), which can be regarded as bosons with
       strong repulsion ($1/\gamma = +0$, $a<0$). 
 \item The above regime of $\gamma$ can be smoothly connected to the
       weakly attractive fermi gas and the weakly repulsive bose gas at
       both ends with sufficiently small value of $|\gamma|$. 
       This reflects the analyticity of the
       Love equation~\eqref{eq.Love} in $\lambda$ on the whole real axis 
       excluding the point $\lambda=0$.    
\end{enumerate}

In the weak coupling case, we have again met the difficulty of
divergences. Obviously, the solution~\eqref{eq.small_F} holds only for 
$x\in [-1+\lambda,1-\lambda]$ (see
Appendix~\ref{ap.another_small_lambda}). Near the edges $|x|\approx 1$,
a different method might be necessary to determine the precise behavior
of the solution. Further, the relation between $\lambda$ and $\gamma$
again is important, since the regularization of $\lambda$ by $\gamma$
yields the finite expressions of the physical quantities. Up to now, our
method is powerful enough to obtain the first two terms of the
asymptotic expressions. We believe that other physical quantities, such
as the dressed charge and the dressed energy, and other quantum
integrable systems can be studied by use of this method. Those subjects
remain as future problems.    

\appendix

\bigskip

\bigskip

\begin{flushleft}
 {\Large\textbf{Appendices}}
\end{flushleft}

\section{Another Derivation of~\eqref{eq.lowest_ap} and 
	\eqref{eq.second_ap}}
\label{ap.another_small_lambda}
In this appendix we present the expansion of the integral, 
\begin{equation}
 J(x,\lambda)\equiv 
  \frac{\lambda}{\pi}\int_{-1}^{1}dy\,\frac{F(y)}{\lambda^2 +(x-y)^2}
  =
  \frac{\lambda}{\pi}\int_{-1}^{1}dy\, 
  \frac{\sum\limits_{m=0}^\infty a_m y^{2m}}
  {\lambda^2 +(x-y)^2},
  \label{intj}
\end{equation}
in power series of $\lambda$ for the case $\lambda \ll 1$. Since 
\begin{equation}
 \frac{1}{\lambda^2
 +(x-y)^2}=
 \begin{cases}
  \displaystyle\frac{1}{(x-y)^2}
  \frac{1}{1+\left(\displaystyle\frac{\lambda}{x-y}\right)^2
  }=
  \displaystyle\sum_{n=0}^{\infty}\displaystyle
  \frac{(-1)^n\lambda^{2n}}{(x-y)^{2n+2}}, 
  & \text{if $|x-y|>\lambda$,}\\
  \displaystyle\frac{1}{\lambda^2}\frac{1}{1+\left(\displaystyle\frac{x-y}{
  \lambda}\right)^2}
  =\displaystyle\sum_{n=0}^{\infty}\frac{(-1)^n(x-y)^{2n}}{\lambda^{2n+2}},
  & \text{if $|x-y|<\lambda$,}\\
 \end{cases}
\end{equation}
for a fixed $x\in [-1+\lambda,1-\lambda]$, \eqref{intj} becomes
\begin{align}
 J(x,\lambda)
 &=\frac{\lambda}{\pi}
 \left[\int_{[-1,x-\lambda)\cup(x+\lambda,1]}+\int_{(x-\lambda,x+\lambda)}
 \right]\frac{dy}
  {\lambda^2 +(x-y)^2}\sum\limits_{m=0}^\infty a_m y^{2m}\nonumber\\
 &=-\frac{2}{\pi}\sum_{m=0}^{\infty} \sum_{n=0}^{\infty} \sum_{l=0}^{m}
 \lambda^{2l}\dbinom{2m}{2l}\frac{(-1)^{n} a_m x^{2(m-l)}}
 {2l-2n-1}\nonumber\\
 &\quad+\frac{1}{\pi}\sum_{m=0}^{\infty} \sum_{n=0}^{\infty} \sum_{k=0}^{2m}
 {}^{'}
 \lambda^{2n+1} \dbinom{2m}{k} \frac{(-1)^n a_m x^{2m-k}}{k-2n-1}\left[
 (-1)^k (1+x)^{k-2n-1}+(1-x)^{k-2n-1}\right]\nonumber\\
 &\quad-\frac{1}{\pi}\sum_{m=1}^{\infty} \sum_{n=0}^{m-1} \lambda^{2n+1}
 \dbinom{2m}{2n+1}(-1)^n a_m x^{2(m-n)-1}\log\frac{1+x}{1-x}\nonumber\\
 &\quad+\frac{2}{\pi} \sum_{m=0}^{\infty} \sum_{n=0}^{\infty} \sum_{l=0}^{m}
 \lambda^{2(m-l)} \dbinom{2m}{2l} \frac{(-1)^n a_m
 x^{2l}}{2n+2m-2l+1}.
\label{anotherexpansion1}
\end{align}
In the above, $\sum\limits_{k=0}^{2m}{}^{'}$ means that we exclude the
$k=2n+1$ terms in the sum with respect to $k$. 

Let us rewrite~\eqref{anotherexpansion1} so that we can easily see the
dependences of $\lambda$ and $x$. For the purpose, we change the
variables and the orders of the multiple sums. To begin with, we
consider the first term of the r.h.s. of~\eqref{anotherexpansion1}. The
order of the triple series can be changed as
\begin{equation}
 \sum_{m=0}^{\infty} \sum_{n=0}^{\infty} \sum_{l=0}^{m}
  =\sum_{l=0}^{\infty}\sum_{m=l}^{\infty}\sum_{n=0}^{\infty}.
\end{equation}
With a new variable $p=m-l$ instead of $m$, the first term of the r.h.s.
of~\eqref{anotherexpansion1} can be written as 
\begin{align}
&-\frac{2}{\pi}\sum_{m=0}^{\infty} \sum_{n=0}^{\infty} \sum_{l=0}^{m}
 \lambda^{2l}\dbinom{2m}{2l}\frac{(-1)^{n} a_m x^{2(m-l)}}
 {2l-2n-1}\nonumber\\
& =-\frac{2}{\pi}\sum_{l=0}^{\infty}\lambda^{2l}\left\{
 \sum_{p=0}^{\infty}\dbinom{2l+2p}{2l}a_{l+p}\left[\sum_{n=0}^{\infty}
 \frac{(-1)^n}{2l-2n-1}\right]x^{2p}\right\}.
\label{first_rhs}
\end{align}
In exactly the same way, the fourth term becomes
\begin{align}
&\frac{2}{\pi} \sum_{m=0}^{\infty} \sum_{n=0}^{\infty} \sum_{l=0}^{m}
 \lambda^{2(m-l)} \dbinom{2m}{2l} \frac{(-1)^n a_m
 x^{2l}}{2n+2m-2l+1}\nonumber\\
& =\frac{2}{\pi}\sum_{p=0}^{\infty}\lambda^{2p}\left\{
 \sum_{l=0}^{\infty}\dbinom{2p+2l}{2p}a_{p+l}\left[
 \sum_{n=0}^{\infty}\frac{(-1)^n}{2n+2p+1}\right]x^{2l}\right\}.
\label{fourth_rhs}
\end{align}
The third term is rewritten as follows: 
\begin{align}
&-\frac{1}{\pi}\sum_{m=1}^{\infty}
 \sum_{n=0}^{m-1}\lambda^{2n+1}\dbinom{2m}{2n+1}
 (-1)^na_mx^{2(m-n)-1}\log\frac{1+x}{1-x}\nonumber\\
 &=-\frac{2}{\pi}\sum_{n=0}^{\infty}\lambda^{2n+1}(-1)^n
 \left[\sum_{m=n+1}^{\infty}\sum_{p=0}^{\infty}\dbinom{2m}{2n+1}\frac{1}{2p+1}
 a_mx^{2(m-n+p)}\right]\nonumber\\
 &=-\frac{2}{\pi}\sum_{n=0}^{\infty}(-1)^n\lambda^{2n+1}
 \left\{\sum_{q=1}^{\infty}\left[\sum_{l=1}^{q}\dbinom{2l+2n}{2n+1}
 \frac{a_{n+l}}{2q-2l+1}\right]x^{2q}\right\}.
\label{third_rhs}
\end{align}
The second term is rather complicated. Since
\begin{align}
 &(-1)^k(1+x)^{k-2n-1}+(1-x)^{k-2n-1}
 =(-1)^{k}\sum_{l=0}^{\infty}\left[1+(-1)^{k-l}\right]
 \dbinom{k-2n-1}{l}x^l\nonumber\\
 &=\begin{cases}
    \displaystyle
    2\sum_{p=0}^{\infty}\dbinom{2n-k+2p}{2p}x^{2p}, & \text{if 
    $0\le k\le 2n$ and $k$ is even,}\\
    \displaystyle
    2\sum_{p=0}^{k/2-n-1}\dbinom{k-2n-1}{2p}x^{2p}, &
    \text{if $2n+2\le k\le 2m$ and $k$ is even,}\\
    \displaystyle
    2\sum_{p=0}^{\infty}\dbinom{2n-k+2p+1}{2p+1}x^{2p+1}, &
    \text{if $1\le k \le 2n-1$ and $k$ is odd,}\\
    \displaystyle
    -2\sum_{p=0}^{(k-1)/2-n-1}\dbinom{k-2n-1}{2p+1}x^{2p+1}, &
    \text{if $2n+3\le k \le 2m-1$ and $k$ is odd,}
  \end{cases}
 \label{2into4}
\end{align}
we decompose the sum with respect to $k$ into four parts:
\begin{equation}
 \sum_{k=0}^{2m}{}^{'}=\sum_{k=0}^{2n}+\sum_{k=2n+2}^{2m}
  =\sum_{\scriptstyle j=0\atop\scriptstyle  k=2j}^{n}+
  \sum_{\scriptstyle j=0\atop\scriptstyle k=2j+1}^{n-1}+
  \sum_{\scriptstyle j=n+1\atop\scriptstyle k=2j}^{m}+
  \sum_{\scriptstyle j=n+1\atop\scriptstyle k=2j+1}^{m-1}.
  \label{kinto4}
\end{equation}
For a fixed $n\ge 0$, the sum with respect to $m$ can be decomposed into
two parts:  
\begin{equation}
 \sum_{m=0}^{\infty}=\sum_{m=0}^{n}+\sum_{m=n+1}^{\infty}.
  \label{minto2}
\end{equation}
With~\eqref{kinto4} and~\eqref{minto2}, the double series with respect
to $m$ and $k$ become
\begin{equation}
 \sum_{m=0}^{\infty}\sum_{k=0}^{2m}{}^{'}
  =\sum_{m=0}^{n}\left(\sum_{\scriptstyle j=0\atop\scriptstyle  k=2j}
		  ^{m}+\sum_{\scriptstyle j=0\atop\scriptstyle k=2j+1}
		  ^{m-1}\right)
  +\sum_{m=n+1}^{\infty}\left(\sum_{\scriptstyle j=0\atop\scriptstyle  k=2j}
			 ^{n}+\sum_{\scriptstyle j=0\atop\scriptstyle k=2j+1}
			 ^{n-1}
  +\sum_{\scriptstyle j=n+1\atop\scriptstyle k=2j}^{m}
  +\sum_{\scriptstyle j=n+1\atop\scriptstyle k=2j+1}^{m-1}\right).
\label{kminto6}
\end{equation}

Now we proceed to evaluate the second term of the
r.h.s. of~\eqref{anotherexpansion1} using~\eqref{kinto4} 
and~\eqref{kminto6}:   
\begin{align}
& \frac{1}{\pi}\sum_{m=0}^{\infty}\sum_{n=0}^{\infty}\sum_{k=0}^{2m}
 {}^{'}
 \lambda^{2n+1}\dbinom{2m}{k}\frac{(-1)^n a_m x^{2m-k}}{k-2n-1}\left[
 (-1)^k (1+x)^{k-2n-1}+(1-x)^{k-2n-1}\right]\nonumber\\
 &=\frac{2}{\pi}\sum_{n=0}^{\infty}(-1)^n\lambda^{2n+1}
 \Biggl\{\sum_{m=0}^{n}a_m
 \sum_{p=0}^{\infty}\Biggl[\sum_{j=0}^{m}\dbinom{2m}{2j}\dbinom{2n-2j+2p}{2p}
 \frac{1}{2j-2n-1}x^{2m-2j+2p}
 \nonumber\\
 &\phantom{
 =\frac{2}{\pi}\sum_{n=0}^{\infty}(-1)^n\lambda^{2n+1}
 \Biggl\{\sum_{m=0}^{n}a_m\sum_{p=0}^{\infty}\Biggl[
 }
 +\sum_{j=0}^{m-1}\dbinom{2m}{2j+1}\dbinom{2n-2j+2p}{2p+1}
 \frac{1}{2j-2n}x^{2m-2j+2p}\Biggr]\nonumber\\
 &
 \quad\quad+\sum_{m=n+1}^{\infty}a_m
 \sum_{p=0}^{\infty}
 \Biggl[
 \sum_{j=0}^{n}\dbinom{2m}{2j}\dbinom{2n-2j+2p}{2p}
 \frac{1}{2j-2n-1}x^{2m-2j+2p}
 \nonumber\\
 &
 \phantom{\quad\quad+\sum_{m=n+1}^{\infty}a_m
 \sum_{p=0}^{\infty}
 \Biggl[}
 +\sum_{j=0}^{n-1}\dbinom{2m}{2j+1}\dbinom{2n-2j+2p}{2p+1}
 \frac{1}{2j-2n}x^{2m-2j+2p}\Biggr]\nonumber\\
 &\quad\quad
 +\sum_{m=n+1}^{\infty}a_m
 \Biggl[
 \sum_{j=n+1}^{m}\sum_{p=0}^{j-n-1}\dbinom{2m}{2j}\dbinom{2j-2n-1}{2p}
 \frac{1}{2j-2n-1}x^{2m-2j+2p}
 \nonumber\\
 &
\phantom{\quad\quad
 +\sum_{m=n+1}^{\infty}a_m
 \Biggl[}
 -\sum_{j=n+1}^{m-1}\sum_{p=0}^{j-n-1}\dbinom{2m}{2j+1}\dbinom{2j-2n}{2p+1}
 \frac{1}{2j-2n}x^{2m-2j+2p}\Biggr]
 \Biggr\}\nonumber\\
 &=
 \frac{2}{\pi}\sum_{n=0}^{\infty}(-1)^n\lambda^{2n+1}
 \left\{\left[\sum_{r=0}^{\infty}\sum_{q=0}^{n}\sum_{l=0}^{2q}
 a_q(-1)^l\dbinom{2q}{l}\dbinom{2q-2n-1-l}{2r-l}
 \frac{1}{2q-2n-1-l}x^{2r}\right.\right.\nonumber\\
 &\phantom{
 (-1)^n
 \lambda^{2n+1}
 \Biggl\{\Biggl[ }
 \quad
 \left.-\sum_{r=0}^{n-1}\sum_{q=r+1}^{n}\sum_{l=2r+1}^{2q}
 a_q(-1)^l\dbinom{2q}{l}\dbinom{2q-2n-1-l}{2r-l}
 \frac{1}{2q-2n-1-l}x^{2r}\right]\nonumber\\
 &\quad \quad+
  \left[
 \sum_{r=1}^{\infty}a_{r+n}\dbinom{2r+2n}{2r}\frac{1}{-1}x^{2r}
 \right.\nonumber\\
 &\quad \quad
 \phantom{-\Biggl[ }
 +\sum_{r=2}^{\infty}\sum_{q=0}^{r-2}\sum_{l=2q+2}^{2r}
 a_{q+n+1}(-1)^l\dbinom{2q+2n+2}{l}\dbinom{2q+1-l}{2r-l}
 \frac{1}{2q+1-l}x^{2r} \nonumber\\
 &\quad \quad
 \phantom{ +\Biggl[ }
 -\sum_{r=n+2}^{\infty}\sum_{q=0}^{r-n-2}\sum_{l=2q+2n+3}^{2r}
 a_{q+n+1}(-1)^l\dbinom{2q+2n+2}{l}\dbinom{2q+1-l}{2r-l}
 \frac{1}{2q+1-l}x^{2r}\Biggr]\nonumber\\
 &\quad
 \quad+\sum_{r=0}^{\infty}\sum_{q=r}^{\infty}\sum_{l=0}^{2r}
 a_{q+n+1}(-1)^l\dbinom{2q+2n+2}{l}\dbinom{2q+1-l}{2r-l}
 \frac{1}{2q+1-l}x^{2r}\Biggr\}.
\label{second_rhs}
\end{align}
To obtain the last expression from the middle, we have again changed the
variables and the orders of the multiple sums. For instance, 
in the first square bracket $[\,\,]$, we have replaced $m,\ j$ and $p$ with 
$q=m$, $l=m-j$, $r=m-j+p$, and used the relation 
\begin{equation}
 \dbinom{n+k-1}{k}=(-1)^k\dbinom{-n}{k}\quad \text{for}\quad n\ge1,\ k\ge 0, 
\end{equation} 
to collect the terms of $2l$ and $2l-1$. 
  
The combination of~\eqref{first_rhs}, \eqref{second_rhs},
\eqref{third_rhs} and~\eqref{fourth_rhs} yields the following form of 
\eqref{anotherexpansion1}:
\begin{align}
 J(x,\lambda)=&\frac{2}{\pi}\sum_{n=0}^{\infty}\lambda^{2n}
 \left[A_{n}(x)+B_{n}(x)\right]
 +\frac{2}{\pi}\sum_{n=0}^{\infty}(-1)^n\lambda^{2n+1}\left[
 C_n(x)+D_n(x)+E_n(x)+F_n(x)\right], 
\label{anotherexpansion2}
\end{align}
where $A_n(x)$, $B_n(x)$, $C_n(x)$, $D_n(x)$, $E_n(x)$ and $F_n(x)$ are
defined as
\begin{align}
 A_n(x)&=
 \sum_{p=0}^{\infty}\sum_{q=0}^{\infty}\dbinom{2n+2p}{2n}
 \frac{(-1)^qa_{n+p}}{2q-2n+1} x^{2p},\\
 B_n(x)&=
 \sum_{p=0}^{\infty}\sum_{q=0}^{\infty}\dbinom{2n+2p}{2n}\frac{(-1)^q
 a_{n+p}}{2q+2n+1}x^{2p},\\
 C_n(x)&=\sum_{p=0}^{n}\left(\sum_{q=0}^{p}\sum_{r=0}^{2q}a_q
 \mathcal{A}_{npqr}x^{2p}
 +\sum_{q=p+1}^{n}
 \sum_{r=0}^{2p}a_q\mathcal{A}_{npqr}x^{2p}\right)
 +\sum_{p=n+1}^{\infty}\sum_{q=0}^{n}\sum_{r=0}^{2q}
 a_q\mathcal{A}_{npqr}x^{2p},\label{eq.c_n}\\
 D_n(x)&=\sum_{p=2}^{\infty}\left(\sum_{q=n+1}^{p-1}\sum_{r=2q-2n}^{2q}\!
 a_q\mathcal{A}_{npqr}x^{2p}+
 \sum_{q=p}^{n+p-1}\!\sum_{r=2q-2n}^{2p}\!a_q \mathcal{A}_{npqr}x^{2p}
 \right)
 +\sum_{p=1}^{\infty}a_{n+p}\mathcal{A}_
 {np,p+n,2p}x^{2p},\\
 E_n(x)&=\sum_{p=0}^{\infty}\sum_{q=n+p+1}^{\infty}\sum_{r=0}^{2p}
 a_q\mathcal{A}_{npqr}x^{2p},\\
 F_n(x)&=
 \sum_{p=1}^{\infty}\sum_{q=1}^{p}\dbinom{2q+2n}{2n+1}\frac{a_{n+q}}{
 2q-2p-1}x^{2p}.\label{eq.f_n}
\end{align} 
Here, the sequence   
\begin{align}
 \mathcal{A}_{npqr}\equiv(-1)^{r}\dbinom{2q}{r}\dbinom{2q-2n-1-r}{2p-r}
 \frac{1}{2q-2n-1-r},\quad n,p,q,r\in\Z_{\ge 0},
 \label{eq.a_npqr}
\end{align}
has been introduced to simplify the expressions. Note that   
sums in the above whose lower bounds are greater than its upper bounds, 
such as $\sum\limits_{q=p+1}^{p}$, are treated as zero.
 
From~\eqref{anotherexpansion2}, we obtain the expansion of
$J(x,\lambda)$ with respect to $\lambda$ as follows:
\begin{equation}
 J(x,\lambda)=\sum_{n=0}^{\infty}\lambda^{n}\mathcal{J}_{n}(x),
\end{equation}
where the first few $\mathcal{J}_n$ are
\begin{align}
 \mathcal{J}_0(x)&=\frac{2}{\pi}\left[A_0(x)+B_0(x)\right],
 \label{j0}\\
 \mathcal{J}_1(x)&=\frac{2}{\pi}\left[C_0(x)+D_0(x)+E_0(x)+F_0(x)\right],
 \label{j1}\\
 \mathcal{J}_2(x)&=\frac{2}{\pi}\left[A_1(x)+B_1(x)\right],
 \label{j2}\\
 \mathcal{J}_3(x)&=-\frac{2}{\pi}\left[C_1(x)+D_1(x)+E_1(x)+F_1(x)\right].
 \label{j3}
\end{align}
From~\eqref{j0}, \eqref{j2} and
\begin{equation}
 \sum_{n=0}^{\infty}\dfrac{(-1)^n}{2n+1}=\dfrac{\pi}{4},
 \label{eq.pi/4}
\end{equation} we easily verify that
\begin{align}
 \mathcal{J}_0(x)=&\sum_{p=0}^{\infty}a_px^{2p}=F(x),\label{eq.J_0}\\
 \mathcal{J}_2(x)=&-\sum_{p=0}^{\infty}(p+1)(2p+1)a_{p+1}x^{2p}.
\end{align}
Use of~\eqref{eq.J_0} in the Gaudin integral
equation~\eqref{eq.dimless_GY} leads to~\eqref{eq.lowest_ap}.  
To simplify the expressions of $\mathcal{J}_{1}(x)$ and
$\mathcal{J}_3(x)$, we need a little calculation and some formulae. The
results are, 
\begin{align}
 \mathcal{J}_1(x)&=\frac{2}{\pi}\sum_{p=0}^{\infty}\left[
 \sum_{n=0}^{\infty}\frac{2p+1}{2n-(2p+1)}
 a_n\right]x^{2p},\label{eq.J_1}\\
 \mathcal{J}_3(x)&=-\frac{2}{\pi}\sum_{p=0}^{\infty}\dbinom{2p+3}
 {3}\left[\sum_{n=0}^{\infty}\frac{a_n}{2n-(2p+3)}\right]x^{2p}.
\end{align} 
Below, we shall explain these shortly. 
We observe that equation~\eqref{eq.J_1} is in accordance
with~\eqref{eq.second_ap}.

Therefore, we conclude that our formal
expansion~\eqref{eq.kernel}
is also valid for small $\lambda$ in the interval $[-1+\lambda,1-\lambda]$. 

\bigskip

\begin{flushleft}
{\it Calculation of $\mathcal{J}_1(x)$}
\end{flushleft}

From~\eqref{j1} with~\eqref{eq.c_n}--\eqref{eq.a_npqr}, we have
\begin{align}
 \frac{\pi}{2}\mathcal{J}_1(x)&=\sum_{p=0}^{\infty}a_0\mathcal{A}_{0p00}x^{2p}
 +\sum_{p=1}^{\infty}\sum_{q=1}^{p}a_q\mathcal{A}_{0pq,2q}x^{2p}
 +\sum_{p=0}^{\infty}\sum_{q=p+1}^{\infty}\sum_{r=0}^{2p}
 a_q\mathcal{A}_{0pqr}x^{2p}\nonumber\\
 &\quad +\sum_{p=1}^{\infty}\sum_{q=1}^{p}\dbinom{2q}{1}
 \frac{a_q}{2q-2p-1}x^{2p}
 \nonumber\\
 &=\sum_{p=0}^{\infty}\frac{1}{-1}a_0x^{2p}+\sum_{p=1}^{\infty}
 \sum_{q=1}^{p}\frac{1}{-1}a_qx^{2p}\nonumber\\
 &\quad +\sum_{p=0}^{\infty}\sum_{q=p+1}^{\infty}a_{q}\left[\sum_{r=0}^{2q}
 (-1)^r\dbinom{2q}{r}\dbinom{2q-1-r}{2p-r}\frac{1}{2q-1-r}\right]x^{2p}
 \nonumber\\
 &\quad +\sum_{p=1}^{\infty}\sum_{q=1}^{p}\frac{2q}{2q-2p-1}a_qx^{2p}
 \nonumber\\
 &=\sum_{p=0}^{\infty}\sum_{q=0}^{p}\frac{1}{-1}a_qx^{2p}
 +\sum_{p=0}^{\infty}\sum_{q=p+1}^{\infty}\frac{2p+1}{2q-(2p+1)}a_qx^{2p}
 +\sum_{p=1}^{\infty}\sum_{q=1}^{p}\frac{2q}{2q-(2p+1)}a_qx^{2p}
 \nonumber\\
 &=\sum_{p=0}^{\infty}\sum_{q=0}^{\infty}\frac{2p+1}{2q-(2p+1)}a_qx^{2p}.
\end{align}
Here, to obtain the third expression from the second, the following
formula has been used:
\begin{align}
  \sum_{l=0}^{p}(-1)^{l}\dbinom{m+1}{l}\dbinom{m-l}{p-l}\frac{1}{m-l}
 =(-1)^p\dbinom{p+1}{1}\frac{1}{m-p},\quad m\ge p+1.
\label{formula_binom1}
\end{align}

\medskip

\begin{flushleft}
{\it Proof of~\eqref{formula_binom1}}
\end{flushleft}

Set
\begin{equation}
 \frac{P_p(m)}{m-p}\equiv
  \sum_{l=0}^{p}(-1)^l\dbinom{m+1}{l}\dbinom{m-l}{p-l}\frac{1}{m-l}.
\label{def_of_P_pm}
\end{equation}
It can be seen that 
\begin{equation}
\text{$P_p(m)$ is a polynomial of $m$ whose degree is equal to or less than
$p$. } 
\label{Psdegree}
\end{equation}
It will be shown that $P_p(m)$ is a constant, and actually, 
\begin{equation}
 P_p(m)=(-1)^p(p+1).
\end{equation}
When $p=0$ or $p=1$, we can verify it directly, so we assume that
$p\ge 2$. 

First, let us denote the $l$-th term of the sum $\sum\limits_{l=0}^{p}$
in the r.h.s. of~\eqref{def_of_P_pm} as $Q_l(m)/(m-p)$, namely, 
\begin{equation}
 P_p(m)=\sum_{l=0}^{p}Q_l(m).
\label{P_and_Q}
\end{equation} 
Then, $Q_l(m)$ is explicitly written as 
\begin{align}
 Q_l(m)&=(-1)^l\dbinom{m+1}{l}\dbinom{m-l}{p-l}\frac{m-p}{m-l}\nonumber\\ 
 &=\begin{cases}\displaystyle
    \frac{(m-1)(m-2)\cdots (m-p)}{p!}, & \text{if $l=0$,}\\
    \displaystyle
    (-1)^l\frac{(m+1)m\cdots \widehat{(m-l+1)}\widehat{(m-l)}
    \cdots (m-p)}{l!\,(p-l)!},  & \text{if $1\le l\le p-1$,}\\
    \displaystyle
    (-1)^p\frac{(m+1)m\cdots (m-p+2)}{p!}, & \text{if $l=p$.}
   \end{cases}
\label{exp_form_Q}
\end{align}
Here, $\cdots\widehat{(\ \,\cdot\ \,)}\cdots$ denotes that such term is
excluded from the sequence of the product.  

From~\eqref{P_and_Q} and~\eqref{exp_form_Q}, $P_p(0),P_p(1),\dots
P_p(p)$ are calculated as follows:
\begin{align}
P_p(k)&=
\begin{cases}
 Q_k(k)+Q_{k+1}(k), & \text{if $0\le k\le p-1$,}  \\
 Q_p(p), & \text{if $k=p$, }
\end{cases}\nonumber\\
&=(-1)^p(p+1).
\label{P_for_0top}
\end{align}
Therefore, \eqref{P_for_0top} with~\eqref{Psdegree} assures that
$P_p(m)=(-1)^p(p+1)\quad \text{for all $m$}$, which completes the proof.  

\bigskip

\begin{flushleft}
{\it Calculation of $\mathcal{J}_3(x)$} 
\end{flushleft}

This could be done in almost the same, but more complicated way than
$\mathcal{J}_1(x)$. We sketch the calculation. From~\eqref{j3}
with~\eqref{eq.c_n}--\eqref{eq.a_npqr}, we have   
\begin{align}
 -\frac{\pi}{2}
 \mathcal{J}_3(x)&=a_0\sum_{p=0}^{\infty}\mathcal{A}_{1p00}x^{2p}+
 a_1\mathcal{A}_{1010}x^{0}+a_1\sum_{p=1}^{\infty}\sum_{r=0}^{2}
 \mathcal{A}_{1p1r}x^{2p}\nonumber\\
 &\quad +\sum_{p=2}^{\infty}\sum_{q=2}^{p}
 \sum_{r=2q-2}^{2q}a_q\mathcal{A}_{1pqr}x^{2p}
 +\sum_{p=1}^{\infty}
 \dbinom{2p+2}{2}\frac{1}{-1}a_{p+1}x^{2p}\nonumber\\
&\quad  +\sum_{p=1}^{\infty}\sum_{q=2}^{p+1}\dbinom{2q}{3}\frac{a_q}{2q-(2p+3)}
 x^{2p}\nonumber\\
 &\quad  +\sum_{p=0}^{\infty}\sum_{q=p+2}^{\infty}a_q
 \left[\sum_{r=0}^{2p}
 (-1)^r\dbinom{2q}{r}\dbinom{2q-3-r}{2p-r}\frac{1}{2q-3-r}\right]x^{2p}
 \nonumber\\
 &=a_0\sum_{p=0}^{\infty}\dbinom{2p+2}{2}\frac{1}{-3}x^{2p}
 +a_1\sum_{p=0}^{\infty}\dbinom{2p+3}{3}\frac{1}{-(2p+1)}x^{2p}
 \nonumber\\
 &\quad +\sum_{p=1}^{\infty}\sum_{q=2}^{p+1}\dbinom{2p+3}{3}
 \frac{a_q}{2q-(2p+3)}x^{2p}
 +\sum_{p=0}^{\infty}\sum_{q=p+2}^{\infty}\dbinom{2p+3}{3}
 \frac{a_q}{2q-(2p+3)}x^{2p}\nonumber\\
 &=\sum_{p=0}^{\infty}\sum_{q=0}^{\infty}
 \dbinom{2p+3}{3}\frac{a_q}{2q-(2p+3)}x^{2p}.
\end{align}  
In the above, the second expression is obtained from the first by
applying the identity,  
\begin{equation}
 \sum_{l=0}^{p}(-1)^l\dbinom{m+3}{l}\dbinom{m-l}{p-l}\frac{1}{m-l}
  =(-1)^p\dbinom{p+3}{3}\frac{1}{m-p},\quad m\ge p+1,
\label{formula_binom2}
\end{equation}
for the sum with respect to $r$ in the last term. 
We note that~\eqref{formula_binom2} can be proved in
a similar way as~\eqref{formula_binom1}. 

\section{Solutions of the Lieb-Liniger and the Gaudin integral
 equations around $1/\lambda=0$}\label{app.infinity}
In this appendix, we solve an integral equation,
\begin{equation}
 H(x,\eta)-\frac{\eta}{\pi}\int_{-1}^{1}\frac{H(y,\eta)\,dy}
  {1+\eta^2(x-y)^2}=\frac{1}{2\pi},\label{eq.general_form_strong}
\end{equation}
around $\eta =0$. We remark that,
$H(x,\eta)$ gives the solutions $F(x)$ of~\eqref{eq.dimless_GY} and $g(x)$
of~\eqref{eq.dimless_LL} in the strong coupling case (i.e. $|1/\lambda|\ll
1$),  
\begin{align}
 F(x)&=4\pi H(x,-1/\lambda),\\
 g(x)&=H(x,1/\lambda).
\end{align}
The physical quantities are expressed as follows, \\
for $\eta\le 0$,
\begin{subequations}
 \begin{align}
  \gamma_{\mathrm{F}}&=\left[4\eta\int_{-1}^{1}
  H(x,\eta)\,dx,\right]^{-1}, \\
  Q&=D_{\mathrm{F}}\left[4\int_{-1}^{1}H(x,\eta)\,dx\right]^{-1},\\
  \frac{E_{\mathrm{F}}^{\mathrm{eff}}}{L}&=
  \frac{D_{\mathrm{F}}^3}{16}\int_{-1}^{1}x^2H(x,\eta)\,dx\cdot
  \left[\int_{-1}^{1}H(x,\eta)\,dx\right]^{-3},
 \end{align}
\end{subequations}
for $\eta\ge 0$,
\begin{subequations}
 \begin{align}
  \gamma_{\mathrm{B}}&=\left[\eta\int_{-1}^{1}H(x,\eta)\,dx\right]^{-1},
  \\
  K&=D_{\mathrm{B}}\left[\int_{-1}^{1}H(x,\eta)\,dx\right]^{-1},\\
  \frac{E_{\mathrm{B}}}{L}&=D_{\mathrm{B}}^3
  \int_{-1}^{1}x^2H(x,\eta)\,dx\cdot
 \left[\int_{-1}^{1}H(x,\eta)\,dx\right]^{-3}.
 \end{align}
\end{subequations}
Hereafter, we denote $H(x,\eta)$ as $H(x)$ for simplicity. 
With the expansions of $H(x)$ and the kernel,
\begin{align}
 H(x)&=\sum_{p=0}^{\infty}b_px^{2p}, \\
 \frac{1}{1+\eta^2(x-y)^2}&=\sum_{n=0}^{\infty}(-1)^n(x-y)^{2n}\eta^{2n},
\end{align}
equation~\eqref{eq.general_form_strong} becomes
\begin{equation}
 \sum_{p=0}b_px^{2p}
  -\frac{2}{\pi}\sum_{l=0}^{\infty}\sum_{n=0}^{\infty}
  \sum_{p=0}^{\infty}(-1)^{n}\eta^{2n+1}\frac{b_p}{2l+2p+1}\dbinom{2n}{2l}
  x^{2(n-l)}=\frac{1}{2\pi},
\end{equation}
which, as in \S~\ref{sec.solutions}, reduces to the following set of
algebraic equations for $\{b_p\}$: 
\begin{equation}
 b_p-\frac{2}{\pi}\sum_{m=0}^{\infty}\sum_{n=p}^{\infty}
  (-1)^{n}\eta^{2n+1}\frac{b_m}{2m+2n-2p+1}\dbinom{2n}{2n-2p}
  =\frac{1}{2\pi}\delta_{p0}. \label{eq.infinity_alg}
\end{equation}
From~\eqref{eq.infinity_alg}, we get
\begin{align}
 b_0&=\frac{1}{2\pi}\biggl[1+\frac{2}{\pi}\eta
 +\frac{4}{\pi^2}\eta^2+\frac{2}{\pi}\left(\frac{4}{\pi^2}
 -\frac{1}{3}\right)\eta^3\nonumber\\
 &\phantom{=\frac{1}{2\pi}\biggl[1
 }-\frac{4}{\pi^2}\left(1-\frac{4}{\pi^2}
 \right)\eta^4-\frac{2}{\pi}\left(\frac{20}{3\pi^2}-\frac{1}{5}
 -\frac{16}{\pi^4}\right)\eta^5\biggr]+O(\eta^6),\\
 b_1&=-\frac{1}{\pi^2}\eta^{3}\left[1+\frac{2}{\pi}\eta-2\left(
 1-\frac{2}{\pi^2}\right)\eta^2\right]+O(\eta^6),\\
 b_2&=\frac{1}{\pi^2}\eta^5+O(\eta^6),\\
 b_p&=O(\eta^{2p+1}),\quad p\ge 3.
\end{align}
By use of the above solution, $\eta$ may be expressed as 
\begin{equation}
 \eta=\pi\xi +\frac{4}{3}\pi^3\xi^4 -\frac{32}{15}\pi^5\xi^6
  +O(\xi^7), 
\end{equation}
where
\begin{equation}
 \xi =\frac{1}{\gamma_{\mathrm{B}}+2}=\frac{1}{4\gamma_{\mathrm{F}}+2}.
\end{equation}
In terms of $\xi$, we have 
\begin{subequations}
 \begin{align}
  b_0&=\frac{1}{2\pi(1-2\xi)}\left(1-\frac{2}{3}\pi^2\xi^3
  \right)+O(\xi^6), \\
  b_1&=-\frac{\pi\xi^3}{1-2\xi}+O(\xi^6),\\
  b_2&=\pi^3\xi^5+O(\xi^6). 
 \end{align}
\end{subequations}
It follows from the above relations that for $\xi \ge 0$, 
\begin{align}
 K&=\pi
 D_{\mathrm{B}}(1-2\xi)\left(1+\frac{4}{3}\pi^3\xi^3\right)+O(\xi^5), \\
 \frac{E_{\mathrm{B}}}{L}&=D_{\mathrm{B}}^3\frac{\pi^2}{3}
 \left(1-2\xi\right)^2\left(1+\frac{32}{15}\pi^2\xi^3\right)+O(\xi^6).
\end{align}
Other physical quantities are readily obtained, which
reproduce~\eqref{eq.strong_fermi} and~\eqref{eq.strong_bose}.   

\section{Proof of~\eqref{newsum1}}
\label{app.pf_sum}
We begin with the following expression of the complete elliptic integral
of the first kind, 
\begin{equation}
 K(k)=\int_{0}^{\frac{\pi}{2}}\frac{d\theta}{\sqrt{1-k^2\sin^2\theta}}
 =\frac{\pi}{2}\sum_{m=0}^{\infty}\left[\frac{(2m-1)!!}{(2m)!!}\right]^2
 k^{2m},\quad |k|<1. 
 \label{elliptic1}
\end{equation}
If we consider the expansion such that 
\begin{equation}
 \frac{2}{\pi}\int_{x}^{1}\frac{dt}{t^{2l+2}}\int_{0}^{\frac{\pi}{2}
 }\frac{d\theta}{\sqrt{1-t^2\sin^2\theta}}
=\sum_{n=-\infty}^{\infty}c_nx^{n},\quad \text{for $x>0$,}
\label{c_expand_of_int}
\end{equation}
we see from~\eqref{elliptic1} that
\begin{equation}
 c_0=\sum_{m=0}^{\infty}\left[\frac{(2m-1)!!}{(2m)!!}\right]^2
 \frac{1}{2m-2l-1}.\label{sofint}
\end{equation}

In the l.h.s. of~\eqref{c_expand_of_int}, since the double definite
integral is convergent for $x>0$, we can invert the order of the
integrations.    
To integrate with respect to $t$, we change the
variable from $t$ to $\varphi$, where $\varphi$ is defined as
$\varphi=\arcsin(t\sin\theta),\ 0\le\varphi\le \pi/2$. 
Then the integration in the l.h.s. of~\eqref{c_expand_of_int} becomes 
\begin{equation}
 \frac{2}{\pi}\int_{x}^{1}\frac{dt}{t^{2l+2}}\int_{0}^{\frac{\pi}{2}}
 \frac{d\theta}{\sqrt{1-t^2\sin^2\theta}}=
 \frac{2}{\pi}\int_{0}^{\frac{\pi}{2}}d\theta
 \int_{\arcsin(x\sin\theta)}^{\theta} 
 d\varphi\,\frac{\sin^{2l+1}\theta}{\sin^{2l+2}\varphi}.\label{sofint2}
\end{equation}
By use of the formula, 
\begin{equation}
 \int\frac{d\varphi}{\sin^{2l+2}\varphi}=-\frac{(2l)!!}{(2l+1)!!}
 \frac{\cos{\varphi}}{\sin^{2l+1}\varphi}\sum_{k=0}^{l}
 \frac{(2l-2k-1)!!}{(2l-2k)!!}\sin^{2k}\varphi,
\end{equation}
the r.h.s. of~\eqref{sofint2} is rewritten as
\begin{multline}
 \frac{2}{\pi}\int_{0}^{\frac{\pi}{2}}d\theta\, 
 \Biggl\{-\frac{(2l)!!}{(2l+1)!!}\Biggl[\cos\theta
 \sum_{k=0}^{l}\frac{(2l-2k-1)!!}{(2l-2k)!!}\sin^{2k}\theta\\
 -\frac{1}{x^{2l+1}}\sqrt{1-x^2\sin^2\theta}\sum_{k=0}^{l}
 \frac{(2l-2k-1)!!}{(2l-2k)!!}x^{2k}\sin^{2k}\theta\Biggr]\Biggr\}.
\end{multline}
Since the second term in the square bracket $[\ ]$ consists of odd
powers of $x$, only the first term contributes to $c_0$, which gives
\begin{equation}
 c_0=-\frac{2}{\pi}\frac{(2l)!!}{(2l+1)!!}
 \sum_{k=0}^{l}\frac{(2l-2k-1)!!}{(2l-2k)!!}\frac{1}{2k+1}.
\end{equation}
Therefore, we obtain
\begin{align}
 \sum_{m=0}^{\infty}\left[\frac{(2m-1)!!}{(2m)!!}\right]^2\frac{1}
 {2m-(2l+1)}
 &=-\frac{2}{\pi}\frac{(2l)!!}{(2l+1)!!}\sum_{k=0}^{l}
 \frac{(2l-2k-1)!!}{(2l-2k)!!}\frac{1}{2k+1}\nonumber\\
 &=-\frac{2}{\pi}\frac{(2l)!!}{(2l+1)!!}\sum_{k=0}^{l}
 \frac{1}{2l-2k+1}\frac{(2k-1)!!}{(2k)!!},\nonumber\\
 &\phantom{=-\frac{2}{\pi}\frac{(2l)!!}{(2l+1)!!}\sum_{k=0}^{l}
 \frac{1}{2l-2k+1}\quad}
 l=0,1,2,\ldots ,
\end{align}
which is nothing but~\eqref{newsum1}.

\end{document}